\documentclass[a4paper,UKenglish,cleveref, autoref, thm-restate]{lipics-v2021}

\pdfoutput=1 
\hideLIPIcs  

\title{Balancing incentives in committee-based blockchains}

\titlerunning{Denial of Profit Attacks} 

\author{Arian Baloochestani}{Department of Electrical Engineering and Computer Science, University of Stavanger, Norway }{arian.masoudbaloochestani@uis.no}{}{}

\author{Leander Jehl}{Department of Electrical Engineering and Computer Science, University of Stavanger, Norway}{leander.jehl@uis.no}{}{}

\authorrunning{A. Baloochestani and L. Jehl} 

\Copyright{Arian Baloochestani and Leander Jehl} 



\category{} 

\relatedversion{} 

\nolinenumbers 

\usepackage{acronym}
\usepackage{xspace}
\usepackage{booktabs}
\usepackage{amsmath}
\usepackage{caption}
\usepackage{array}

\usepackage{tikz}
\usepackage{amssymb}
\usepackage{tikzsymbols}
\usetikzlibrary{shapes}
\usetikzlibrary{shapes.misc, positioning}

\acrodef{PoS}{Proof-of-Stake}
\acrodef{PoW}{Proof-of-Work}

\newcommand{\cost}{\textit{cost}\xspace}

\newcommand{\Cost}{\textit{Cost}\xspace}

\newcommand{\effectiveness}{\textit{effectiveness}\xspace}
\newcommand{\Effectiveness}{Effectiveness\xspace}
\newcommand{\eff}{\textnormal{effectiveness}\xspace}
\newcommand{\co}{\textnormal{cost}\xspace}
\newcommand{\power}{\textnormal{P}\xspace}
\newcommand{\pow}[1][i]{\ensuremath{\textnormal{P}[#1]}\xspace}

\newcommand{\dop}{{Denial of Profit}\xspace}

\newcommand{\Basereward}{\textit{Base reward}\xspace}
\newcommand{\basereward}{\textit{base reward}\xspace}

\newcommand{\Aggregation}{\textit{Aggregation}\xspace}
\newcommand{\aggregation}{\textit{aggregation}\xspace}

\newcommand{\Window}{\textit{Inclusion window}\xspace}
\newcommand{\window}{\textit{inclusion window}\xspace}

\newcommand{\Scaling}{\textit{Scaling rewards}\xspace}
\newcommand{\scaling}{\textit{scaling rewards}\xspace}

\newcommand{\Threshold}{\textit{Bonus threshold}\xspace}
\newcommand{\threshold}{\textit{bonus threshold}\xspace}

\newcommand{\arrowup}{\tikz[baseline=-0.5ex] \draw[->, thick, color=black, line width=0.7mm] (0,0) -- (0,0.5);}
\newcommand{\arrowdown}{\tikz[baseline=-0.5ex] \draw[->, thick, color=black, line width=0.7mm] (0,0.5) -- (0,0);}

\newcommand{\powersum}{\ensuremath{\Sigma\power}\xspace}

\newboolean{showcomments}
\setboolean{showcomments}{false}
\ifthenelse{\boolean{showcomments}}
{ \newcommand{\mynote}[3]{
		\textcolor{#3}{{\bfseries\sffamily\scriptsize#1: }\small#2}
}}
{ \newcommand{\mynote}[3]{}}

\begin{document}

\maketitle

\begin{abstract}
Blockchain protocols incentivize participation through monetary rewards, assuming rational actors behave honestly to maximize their gains. 
However, attackers may attempt to harm others even at personal cost.
These \dop attacks aim to reduce the rewards of honest participants, potentially forcing them out of the system.
While existing work has largely focused on the profitability of attacks, they often neglect the potential harm inflicted on the victim, which can be significant even when the attacker gains little or nothing.

This paper introduces a framework to quantify \dop attacks by measuring both attacker cost and victim loss. 
We model these attacks as a game and introduce relevant metrics to quantify these attacks.
We then focus on committee-based blockchains and model vote collection as a game.
We show that in the vote collection game, disincentivizing one \dop attack will make another attack more appealing, and therefore, attacks have to be balanced.
We apply our framework to analyze real-world reward mechanisms in Ethereum and Cosmos. 
Our framework reveals imbalances in Cosmos that can make correct behavior suboptimal in practice. 
While Ethereum provides stronger protections, our framework shows that it is also not complete, and we propose alternative parameter settings to improve the balance between attacks.
Our findings highlight the need for better-balanced reward designs to defend against \dop attacks.
\end{abstract}

\section{Introduction}
\label{sec:intro}
Blockchain technology promises decentralized trust and secure record-keeping. 
To provide a decentralized alternative to centralized cloud providers, blockchains rely on attracting a diverse set of contributors by providing monetary rewards.
However, due to the pseudonymous nature of these systems, they also attract doubtful participants who may try to harm the system or its participants.

Attacks may not always result in financial gain, and the motivations of attackers are diverse.
For example, denial-of-service attacks on SmartContracts may render a contact dysfunctional and funds locked within the contract without direct gain for the attacker~\cite{mastering_ethereum}.
The well-known selfish mining attack focuses on reducing others' rewards rather than increasing one's own~\cite{selfishmining}. Recent work has shown evidence of the attack in practice~\cite{selfishdetection}.
Similarly, an attacker may infiltrate a mining pool to reduce the pool's profitability~\cite{blockwithholding} and thus significantly affect pool populations~\cite{miningpoolselection}.

Despite these examples, the analysis of rewarding systems is often limited to game theoretic equilibria and the profitability of attacks~\cite{amoussou2024committee, amoussou2019fairness, fooladgar2020incentive, motepalli2021reward, sapirshtein2016optimal, wu2019equilibrium}. 
Such analysis is important and may establish trust in the overall system. 
The rationale behind such analysis is that attacks performed at a loss are unlikely to reach a scale that is harmful to the system.
However, we argue that such analysis is insufficient to provide guarantees for the reward received by an honest participant. 
While participants in blockchain systems may act through one or multiple pseudonyms, many key participants, such as mining or staking pools, opt to reveal their identity to make their level of contribution to the ecosystem, as well as their profitability, transparent. 
Known contributors may gain trust and reputation among clients, open source contributors, or participants in governance processes.
However, this duality between known and pseudonymous participants favors targeted attacks, which may significantly harm the reputation of known participants.
Ultimately, the victim of such a targeted attack may even leave the system, seeking profit elsewhere. 
In the long run, such \textit{\dop} attacks may be profitable for the attacker, even if the attack itself is performed at a loss~\cite{buterin2018discouragement}.

We propose a model and metrics to measure the impact that targeted \dop attacks can have on the rewards received by a well-behaved node.
Different from previous work that considers the effect of arbitrary or byzantine behavior~\cite{amoussou2019rationals, reynouard2024bar, zappala2021game}, we focus on the effect targeted attacks have on individual participants rather than on the overall system.
We also consider the loss in reward this deviation causes to the attacker.

The analysis of such targeted attacks is especially relevant in committee-based blockchains, where rewards are based on cooperation rather than individual contributions.
Committee-based blockchains such as Ethereum~\cite{buterin2018ethereum}, Cosmos~\cite{kwon2016cosmos}, and Algorand~\cite{chen2019algorand} use a designated committee to vote for blocks and thus reach consensus on the state of the ledger. This allows faster finality of blocks than traditional proof-of-work blockchains, given that committee participants provide timely votes for blocks.
To incentivize such timely votes, participants may receive a reward for their vote, but only when this vote is included in a block. 
Similarly, block creators may receive rewards for including votes.
Thus, voters and block creators depend on mutual collaboration to receive rewards and may be attacked by unilateral denial of collaboration.
Denial of collaboration between individual participants may have little impact on the overall system but may significantly reduce the participant's rewards.

We define a normal-form game that models vote collection and investigates a wide range of rewarding mechanisms present in real systems like Ethereum or Cosmos.
We show that reward mechanisms need to balance rewards between voters and vote collectors (block creators) to balance the attacks of unilateral denial of collaboration by one of these parties. 
In particular, we show that adjusting the rewarding function to render one of these attacks less effective or more costly will render the other attack more attractive, i.e., more effective or less costly.
This result reflects the nature of collaboration between two parties.
We also show that this limitation can be circumvented by introducing additional parties to the collaboration, e.g., separating vote collection and block creation.

Finally, we evaluate the actual rewarding function and parameters used in two blockchain systems, namely Cosmos and Ethereum.
Our analysis shows that ignoring the balance renders good behavior a non-optimal strategy in Cosmos. 
Ethereum is the system achieving the best balance in our analysis.
However, the rewards are not completely balanced. 
For example, an attacker with 15\% of the stake can deny any victim up to 3\% of their total reward. 
If the stakes are equal, the attacker only loses 2.1\% of their reward in this attack. 

In summary, the contributions of the paper are as follows: 
\begin{description}
    \item[Quantification framework] We develop a framework that quantifies the effectiveness of Denial of Profit attacks.
    The developed framework is general enough to be applied to different protocols.
    The quantification includes both the maximum loss inflicted on the victim and the cost the attack demands from the attacker.
    Additionally, the quantification encompasses that loss and cost may depend on the attacker's and victims' power or stake in the system.
    \item[Vote collection game] We define a normal form game modeling vote collection and show that to avoid favoring one of the parties (voters or collectors), the cost and effectiveness of collaboration denial have to be balanced.
    \item[Mechanism evaluation] We extract several protection mechanisms from systems in production and evaluate their effectiveness using our framework. Our framework helps to better understand and categorize different protection mechanisms.
    \item[Evaluation on validator rewards] We apply our framework to quantify the level of protection against discouragement attacks in the Cosmos and Ethereum blockchains and propose parameters that better balance attacks.
\end{description}

\section{Background}
\label{sec:background}

A blockchain is an append-only ledger, stored and maintained by multiple distributed participants.
Each block contains data, such as transactions, and a cryptographic hash linking it to the previous block.
This structure ensures data integrity and enables participants to maintain a shared log without having a central authority. 
The position of a block, starting from the initial genesis block is called the block height.
To preserve consistency, participants run a consensus algorithm to agree on the order in which blocks are added.

The most well-known consensus mechanism is Proof of Work (PoW), adopted by Bitcoin.
In PoW, participants (or miners) compete to solve a computationally hard puzzle, and the first to succeed earns the right to append a new block.
While PoW provides strong security under certain assumptions, it suffers from high energy consumption, limited throughput, and the possibility of creating forks, situations where multiple valid blocks are produced at the same height, causing temporary disagreement about the state of the ledger.

To address PoW challenges, Proof-of-Stake (PoS) protocols were introduced.
In PoS, participants are selected to create new blocks based on the amount of cryptocurrency they hold and are willing to "stake" as collateral.
This reduces energy consumption and can increase performance, but still requires mechanisms to coordinate among participants and avoid forks.

One such mechanism is to assign a set of participants to validate new blocks and resolve conflicts between proposed blocks by running a consensus algorithm. 
These validators need to have "staked" collateral and run a node participating in the consensus algorithm.
The blockchains adopting this mechanism are called committee-based blockchains, since the committee of validators jointly creates and finalizes new blocks.
At each block height, a leader is chosen to propose the next block, while the rest of the committee votes to accept and commit the proposed block.
A block is committed once enough votes are collected, typically reaching a predefined threshold such as $2/3$. 
Committee-based consensus protocols have recently gained popularity for their ability to provide fast finality, low latency, and predictable throughput, while significantly reducing computational overhead.
Well-known examples include Ethereum~\cite{buterin2018ethereum}, Algorand~\cite{chen2019algorand}, Tendermint (Cosmos)~\cite{kwon2014tendermint, kwon2016cosmos}, Dfinity~\cite{hanke2018dfinity}, Snow White~\cite{daian2019snow}, and Harmony~\cite{harmony} each using different methods for committee and leader selection.

The open and pseudonymous nature of blockchain participation introduces significant challenges.
Blockchains host a diverse set of participants, ranging from anonymous individuals to well-known public entities and institutional validators representing thousands of small investors.
While many participants contribute honestly to maintain the system’s integrity, others may behave adversarially, attacking one another or even the blockchain itself, despite the potential damage to its reputation.
Therefore, many blockchains design reward systems to incentivize desired behavior and ensure that participants act in accordance with the protocol.
These rewards are typically tied to validators' contributions, such as timely voting, block proposal, and up-time.

\section{Metrics for Denial of Profit Attacks}
\label{sec:game}
In \dop attacks, an attacker tries to reduce the reward or profit received by one or multiple victims.
We are especially interested in the case where both the attacker and victim are participating in some game where they are rewarded for their correct participation, e.g., both are running staked validator nodes.

The typical strategy for a \dop attack is for the attacker to try and censor the victim's contributions.
We note that, total censorship is not needed, since already the censoring of an individual contribution may affect the victims reward. 
Censoring the victim's contribution may also reduce his reputation as a reliable supporter of the protocol.

We assume the attacker is willing to perform the attack even at personal cost.
However, the cost an attack causes the attacker is clearly a relevant factor, and we assume that attacks that incur significant costs for the attacker but only have minor effects on the victim are unlikely to happen.
Thus, to establish the danger of a specific \dop attack and the robustness of a protocol against such attacks, we need to evaluate both how much harm can be done by such attacks, and what is the cost for the attacker.

We define \effectiveness and \cost metrics for attacks on a designated strategy in a normal form game below.
Our definitions also include a power function, that allows to model games where different participants have larger or smaller influence.
Power allows both to model coalitions or to model the deposited stake in the reward systems in committee-based blockchains.

\subsection{Game Model}

Let 
$G = \langle N, S, U \rangle$, be a normal form game, where $N=\{p_1, ..., p_n\}$ is the player set, $S$ is the strategy set, and $U$ is the utility function. 

We assume that a designated nash-equilibrium strategy 
profile for $G$ exists, denoted $s_e$.
Finally, we assume that there exists a power function $\pow[\cdot]: N \rightarrow [0,0.33]$ that assigns every player $p_i\in N$ a power value $\pow[i]$. 
We assume that in $s_e$ the utilities of players are relative to their power, i.e. $U(p_i,s_e) = \frac{\pow[i]}{\pow[j]}\cdot U(p_j,s_e)$.

In the case of a distributed protocol the strategy profile $s_e$ describes following the protocol rules and the power function $\pow$ represents the stake distribution or the voting power of the players.

\subsection{Attack metrics}

An attack is a strategy profile $s_a$ where all but one player follow $s_e$. 
Let $p_a$ denote the player not following $s_e$ in $s_a$.

Since $s_e$ is a nash-equilibrium, $U(p_a, s_a)\leq U(p_a,s_e)$ holds.
We are interested in how the deviation by the attacker effects the utility of the other players.
We define \effectiveness to measure the damage done to the other players by the attacker and \cost to measure the cost of the attack to the attacker.

\begin{definition}
We define the \effectiveness of the attack $s_a$ as the maximum loss in utility done to other players, divided by the power of the attacker.

\begin{equation}
    \label{eq:eff}
    \eff(s_a) = \max_{p\in N, p \neq p_a} \frac{U(p,s_e) - U(p,s_a)}{U(p,s_e) \cdot \pow[a]}
\end{equation}
\end{definition}

\Effectiveness measures the damage an attack does to an individual player, rather than the effect an attack has on the overall system. 
Thus, \effectiveness is especially useful, to analyze targeted attacks on individual players.
We note that in practice, a more powerful player will be able to do more harm to others than a less powerful player. 
However, since open blockchain often allow individual operators to control different pseudonymous identities, it is difficult to determine the potential power of an attacker. Further, an operator may also decide to perform an attack only with a fraction of their power. 
Therefore, we normalize the damage done by the power of the attacker. This makes \effectiveness a useful metric to compare the damage done by different attacks, independent of the power of the attacker.
In many of the cases we analyse, this makes the \effectiveness independent of the power distribution.

In games without negative utility, \effectiveness will usually be positive, unless all other players benefit from the attack. 
An effectiveness of $0$ means that the attack does not harm any other player. With an effectiveness of $1$, the effect of the attack is relative to the attackers power. Thus an attacker with $10\%$ power, can reduce another players utility by $10\%$.
Finally, an effectiveness larger than $1$ means that that attacker can do significant harm. 
For example, if an attacker with power 1/3 would manage to reduce the utility of another player to zero, the \effectiveness is 3.
If an attacker with $10\%$ power manages the same, the 
\effectiveness is 10.

\begin{definition}
    We define the \co of the attack $s_a$ as the loss of the attacker, relative to the maximum loss of another players:
    \begin{equation}
        \co(s_a) =\frac{U(p_a,s_e) - U(p_a,s_a)}{
            \max_{p\in N, p\neq p_a} (U(p,s_e) - U(p,s_a))}
    \end{equation}
\end{definition}

Again, \co is designed as a metric for targeted attacks, comparing the attackers loss to the loss of the victim affected most. The metric does not consider wether other players also loose or gain utility.

A \co larger than 1 means that the attacker looses more than the most affected victim. On the other hand, a \co smaller than 1 is quite effective, since the victims utility is reduced more than the attackers.
Finally, a cost below 0 means that the attacker gains utility from the attack.
This implies that the designated strategy is no nash-equilibrium.

\subsection{Robustness}
The metrics of \effectiveness and \co apply to individual attacks. 
We can use these metrics to define the robustness of an equilibrium, by considering the worst case attack.

\begin{definition}
For to parameters $\epsilon, \gamma \geq 0$, a strategy profile $s_e$ is $\epsilon-\gamma$-robust, 
if for all attacks $s_a$ on $s_e$, $\eff(s_a)\leq\epsilon$ and $\co(s_a)\geq\gamma$ holds.
\end{definition}

In the context of decentralized protocols, robustness is a measure useful to determine the security of a protocol and the risk of participation. 
In this context, nash-equilibria have traditionally been used as a metric for system designers, specifying that participants are likely to follow a given protocol.
If the protocol is for example a $0.1-2$-robust equilibrium, a participant can be safe that the deviation of individual participants with less than $10\%$ power cannot reduce their utility by more than $1\%$. 
Further, for every token the participant does not receive, due to an attack, the attacker looses at least $2$ tokens.
Thus, from the view of participants, robustness is a valuable property.

\section{Vote Collection Game}
\label{sec:vote}
In this section, we define a normal form game that models vote collection in committee-based blockchain systems. 
We then show how our attack metrics can be applied to vote omission and vote delay attacks in this game.
We also introduce different mechanisms that existing systems use to protect against these attacks and analyze their effect on \cost and \eff.

In committee-based proof-of-stake blockchains, validators repeatedly create and sign blocks.
Validators are nodes that have deposited a certain amount of cryptocurrency as a stake.
In every round, a leader collects signatures from the validators and includes them in the next block.
The leader is chosen at random or according to a predefined schedule.
Validators, including the leader, are rewarded based on the signatures included in the block.
We model a single round of this protocol as a normal form game.

We assume a set of $n$ players $N = \{p_1, p_2, \ldots, p_n\}$, each of whom control one or more validators. 
The power function $\power: N \rightarrow [0, 1/3]$ models the fraction of the stake a player controls in the system.
If a player controls multiple validators, the player's power is the sum of the stakes of the validators.
We assume that no player controls more than 1/3 of the total stake in the system and that the power of all players sums up to 1.

We model the actions a player can take, when they control the leader or their validators as a strategy set containing strategy $s_e$ and for every player $p_i$, strategies $s_i^l$, $s_i^v$, and $s_i$.
$S= \{s_e, s_1^l, s_1^v, s_1, s_2^l, ...\}$.
Here, strategy $s_e$ is the designated strategy, where the leader includes all signatures in the block and players always let their validators submit a signature.
Strategies $s_i^{(l/v)}$ are designed to attack player $p_i$. 
Strategy $s_i^l$ models a vote omission attack against player $p_i$.
If a player following $s_i^l$  is assigned the leader role, it omits the signature of player $p_i$, otherwise, it follows $s_e$. 
Strategy $s_i^v$ models a vote delay attack.
A validator from a player following $s_i^v$ does omit or delay its signature, when $p_i$ is the leader, and otherwise follows $s_e$. 
Finally, a player following $s_i$ combines both attacks, following $s_i^l$, when it is the leader and $s_i^v$ otherwise.
We write $S_e$ for the strategy profile, where all players follow the designated strategy $s_e$ and $S_{j \rightarrow i}^{(l/v)}$ for the strategy profile, where all players follow the designated strategy, except the player $p_j$ who follows the strategy $s_i^{(l/v)}$. (Player $p_j$ attacks $p_i$ in $S_{j \rightarrow i}^{(l/v)}$)

The utility function gives the expected reward a player's validators receive for a round.
We assume that the reward a player $p_i$ receives can be expressed as a function $R(\delta_l,\delta_i, \pow[i], \powersum)$, where 
$\delta_l\in\{0,1\}$ expresses if $p_i$ is the leader of the round, $\delta_i\in\{0,1\}$ expresses if $p_i$'s signatures are included in the next block, $\pow[i]$ is the power of player $p_i$, and $\powersum$ is the total power of signatures included by the leader.

The reward for a round is probabilistic, since the leader is elected at random. 
Let $P_l(p_i)$ be the probability, that $p_i$ is elected as leader.
Thus, in the designated strategy $S_e$, each player $p_i$ has utility shown in Equation~(\ref{eq:utilSe}), where we write $P_l^c(p_i)$ for $1-P_l(p_i)$:
\begin{equation}
    \label{eq:utilSe}
    U(p_i,S_e)= P_l(p_i)\cdot R(1,1,\pow[i],1)+ P_l^c(p_i)\cdot R(0,1,\pow[i],1)   
\end{equation}
For strategy profile $S^l_{j \rightarrow i}$, the utility of the attacker $p_j$ is given by Equation~(\ref{eqU:lj}).
\begin{equation}
    \label{eqU:lj} 
    U(p_j,S^l_{j \rightarrow i})
    = P_l(p_j) \cdot R(1,1,\pow[j],1-\pow[i])
    + P_l^c(p_j) \cdot R(0,1,\pow[j],1)
\end{equation}
The utility of the victim $p_i$ is given in Equation~\ref{eqU:li}.
The terms in the sum contain the cases that $p_i$ is the leader, that $p_j$ is the leader and omits $p_i$, and that none of $p_i$ or $p_j$ are the leader.
\begin{equation}
    \label{eqU:li} 
    \begin{split}
    U(p_i,S^l_{j \rightarrow i})= & P_l(p_i) \cdot R(1,1,\pow[i],1)\\
    + & P_l(p_j) \cdot R(0,0,\pow[i],1-\pow[i]) \\
    + & (1- P_l(p_j) - P_l(p_i)) \cdot R(0,1,\pow[i],1)
    \end{split}
\end{equation}

Utility for other players $U(p_k ,S^l_{j \rightarrow i})$, 
and utilities for $S^v_{j \rightarrow i}$ can be derived similarly.
For the rest of this paper, we assume that the chance for a player to be elected as leader is proportional to its power, i.e. $P_l(p_i)=\pow[i]$.

\begin{figure}
    \centering
    \subfloat[Rewarding scheme in absence of attacks where voters (validators) receive rewards and block creator (proposer) receives his bonus.]{        \begin{tikzpicture}[node distance=0cm,scale=0.9]

            \colorlet{color1}{cyan}
            \colorlet{color2}{green}
            \colorlet{color3}{yellow!70!gray}
            \colorlet{color4}{orange}
            \colorlet{color5}{magenta}
            \colorlet{colorx}{gray}
            
            \newcommand{\emojihead}[3]{%
              \node at (#1, -2) {\textcolor{#2}{\scalebox{0.8}{#3}}};
            }
            \newcommand{\cancel}[2]{%
              \node[draw, cross out, ultra thick] at (#1, #2) {};
            }
            
            \emojihead{0}{color1}{\Smiley[3]}
            \emojihead{1}{color2}{\Smiley[3]}
            \emojihead{2}{color3}{\Smiley[3]}
            \emojihead{3}{color4}{\Smiley[3]}
            \emojihead{4}{color5}{\Smiley[3]}
            
            \node (prop) at (-1, 1) {\scalebox{0.8}{\textcolor{black}{\Smiley[3]}}};
            \node[single arrow, draw, minimum height=4.4cm, minimum width=1cm,shape border rotate=180,
                single arrow head extend=0.15cm,
            ] (arr) [right=of prop] {};
            
            \newcommand{\letter}[4][0]{%
              \node at (#2, #1) {\textcolor{#3}{\huge $\mathcal{#4}$}};
            };

            \newcommand{\letters}[4]{%
              \node at (#1, #2) {\textcolor{#3}{$\mathcal{#4}$}};
            };

            \letter{0}{color1}{A};
            \letter{1}{color2}{B};
            \letter{2}{color3}{C};
            \letter{3}{color4}{D};
            \letter{4}{color5}{E};




            \draw[ultra thick] (-0.5, -0.5) rectangle (4.5, 0.5);
            \draw[ultra thick] (-1.5, -0.5) rectangle (-1, 0.5);
            \draw[ultra thick] (-1, 0) -- (-0.5, 0);
            \draw[color=white,fill=white] (-1.6, -0.6) rectangle (-1.4, 0.6);
            

            \newcommand{\coloredcircle}[4][1]{%
                \node[draw, circle, fill=#2, minimum size=0.2cm, inner sep=0.02cm] at (#3, #4) {\scalebox{#1}{\textcolor{black}{\textdollar}}};
            }

            \newcommand{\coloredcirclex}[4][1]{%
                \node[draw, circle, fill=#2, minimum size=0.2cm, inner sep=0.02cm] at (#3, #4) {\textcolor{black}{\textdollar}};
                \node[fill=white, circular sector,circular sector angle=72,rotate=-54, inner sep=0.07cm] (cp) [left={0cm of (#3, #4)}] {};
            }
            \newcommand{\coloredcircley}[4][1]{%
                \node[draw, circle, fill=#2, minimum size=0.2cm, inner sep=0.02cm] at (#3, #4) {\scalebox{0.6}{\textcolor{black}{\textdollar}}};
                \node[fill=white, circular sector,circular sector angle=72,rotate=-54, inner sep=0.045cm] (cp) [left={0cm of (#3, #4)}] {};
            }
            \newcommand{\coloredcirclez}[4][1]{%
                \node[draw, circle, fill=#2, minimum size=0.2cm, inner sep=0.02cm] at (#3, #4) {\textcolor{black}{\textdollar}};
                \node[fill=white, circular sector,circular sector angle=90,rotate=45, inner sep=0.08cm] (cp) [left={0cm of (#3, #4)}] {};
            }

            \foreach \x/\col in {0/color1, 1/color2, 2/color3, 3/color4, 4/color5} {
              \draw[->, thick] (\x, -0.3) -- (\x, -1.5);
              \coloredcircle{\col}{\x - 0.2}{-1};
            };


            \foreach \x/\col in {0/color1, 1/color2, 2/color3, 3/color4} {
                \coloredcircle[0.6]{\col}{\x}{1};
            };
            \coloredcircle[0.6]{color5}{4}{1};
            \node at (2,-2.7) {validators};
            \node [above=-0.2cm of prop] {proposer};
            
        \end{tikzpicture}}
    \hfill
    \subfloat[Reward examle: Vote $\mathcal{E}$ was omitted or delayed and the reward for validator $\mathcal{E}$ and the proposer bonus are reduced. 
        ]{        \begin{tikzpicture}[node distance=0cm,scale=0.9]

            \colorlet{color1}{cyan}
            \colorlet{color2}{green}
            \colorlet{color3}{yellow!70!gray}
            \colorlet{color4}{orange}
            \colorlet{color5}{magenta}
            \colorlet{colorx}{gray}
            
            \newcommand{\emojihead}[3]{%
              \node at (#1, -2) {\textcolor{#2}{\scalebox{0.8}{#3}}};
            }
            \newcommand{\cancel}[2]{%
              \node[draw, cross out, ultra thick] at (#1, #2) {};
            }
            
            \emojihead{0}{color1}{\Smiley[3]}
            \emojihead{1}{color2}{\Smiley[3]}
            \emojihead{2}{color3}{\Smiley[3]}
            \emojihead{3}{color4}{\Smiley[3]}
            \emojihead{4}{color5}{\Sadey[3]}
            
            \node (prop) at (-1, 1) {\scalebox{0.8}{\textcolor{black}{\Neutrey[3]}}};
            \node[single arrow, draw, minimum height=4.4cm, minimum width=1cm,shape border rotate=180,
                single arrow head extend=0.15cm,
            ] (arr) [right=of prop] {};
            
            \newcommand{\letter}[4][0]{%
              \node at (#2, #1) {\textcolor{#3}{\huge $\mathcal{#4}$}};
            };

            \newcommand{\letters}[4]{%
              \node at (#1, #2) {\textcolor{#3}{$\mathcal{#4}$}};
            };

            \letter{0}{color1}{A};
            \letter{1}{color2}{B};
            \letter{2}{color3}{C};
            \letter{3}{color4}{D};
            \letter{4}{colorx}{E};
            \cancel{4}{0};




            \draw[ultra thick] (-0.5, -0.5) rectangle (4.5, 0.5);
            \draw[ultra thick] (-1.5, -0.5) rectangle (-1, 0.5);
            \draw[ultra thick] (-1, 0) -- (-0.5, 0);
            \draw[color=white,fill=white] (-1.6, -0.6) rectangle (-1.4, 0.6);
            

            \newcommand{\coloredcircle}[4][1]{%
                \node[draw, circle, fill=#2, minimum size=0.2cm, inner sep=0.02cm] at (#3, #4) {\scalebox{#1}{\textcolor{black}{\textdollar}}};
            }

            \newcommand{\coloredcirclex}[4][1]{%
                \node[draw, circle, fill=#2, minimum size=0.2cm, inner sep=0.02cm] at (#3, #4) {\textcolor{black}{\textdollar}};
                \node[fill=white, circular sector,circular sector angle=72,rotate=-54, inner sep=0.07cm] (cp) [left={0cm of (#3, #4)}] {};
            }
            \newcommand{\coloredcircley}[4][1]{%
                \node[draw, circle, fill=#2, minimum size=0.2cm, inner sep=0.02cm] at (#3, #4) {\scalebox{0.6}{\textcolor{black}{\textdollar}}};
                \node[fill=white, circular sector,circular sector angle=72,rotate=-54, inner sep=0.045cm] (cp) [left={0cm of (#3, #4)}] {};
            }
            \newcommand{\coloredcirclez}[4][1]{%
                \node[draw, circle, fill=#2, minimum size=0.2cm, inner sep=0.02cm] at (#3, #4) {\textcolor{black}{\textdollar}};
                \node[fill=white, circular sector,circular sector angle=90,rotate=45, inner sep=0.08cm] (cp) [left={0cm of (#3, #4)}] {};
            }

            \foreach \x/\col in {0/color1, 1/color2, 2/color3, 3/color4} {
              \draw[->, thick] (\x, -0.3) -- (\x, -1.5);
              \coloredcircle{\col}{\x - 0.2}{-1};
            };


            \foreach \x/\col in {0/color1, 1/color2, 2/color3, 3/color4} {
                \coloredcircle[0.6]{\col}{\x}{1};
            };
            \coloredcircle[0.6]{color5}{4}{1};
            \cancel{4}{1};
            \node at (2,-2.7) {validators};
            \node [above=-0.2cm of prop] {proposer};
            
        \end{tikzpicture}}
    \caption{Reward distribution in Example~\ref{exampleR} with and without attacks.} 
    \label{fig:base-rewarding}
\end{figure}
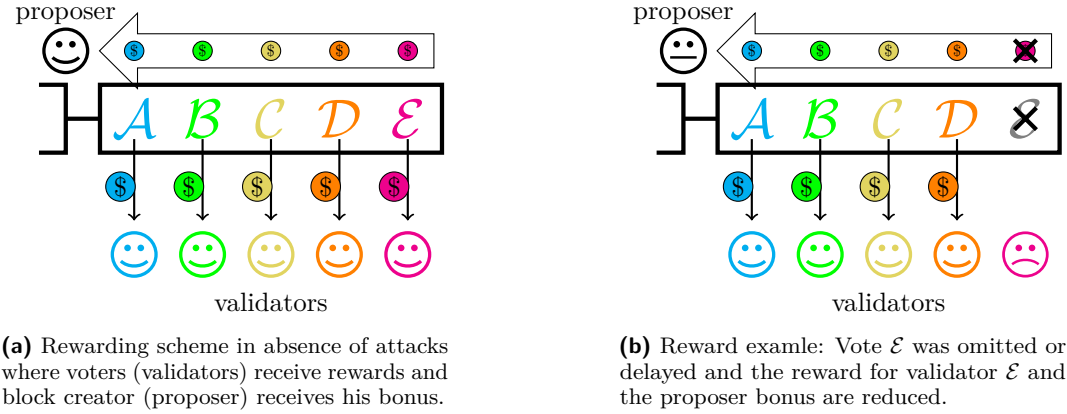

\begin{example}
    \label{exampleR}
Typically, the non-leaders receive a reward proportional to their power, if they are included, while the leader receives an additional reward or bonus proportional to the total power of included signature $\powersum$.
For example, in a simple scheme there may exist parameters $b$ and $R$, such that 
\begin{equation}
    R(\delta_l,\delta_i, \pow[i], \powersum) = \delta_l \cdot \powersum \cdot bR + \delta_i \cdot \pow[i] \cdot R
\end{equation}
Here $R$ is the maximum reward payed out to validators, while the leader receives 
an additional bonus of up to $bR$.
In the designated strategy profile, a player $p_i$ thus receives utility $U(p_i,S_e)= \pow[i] R + \pow[i] bR= \pow[i](1+b)R$.
Figure~\ref{fig:base-rewarding} illustrates the rewarding with and without \dop attacks in this example. 
\end{example}

\subsection{Cost and balance of attacks}

In the vote collection game, as defined above, there exist some interesting relations between the \cost and \effectiveness of the vote omission and vote delay attacks.
Especially, we show that when designing the reward function, one has to balance the vote omission and vote delay attacks. 
Strongly discouraging one attack must necessarily make the other attack more attractive.

We note that \effectiveness and \cost are defined using the maximum loss over all non-attacking players. 
For the targeted attacks $S_{j\rightarrow i}^{(l/v)}$ and all reward functions considered in this paper, the targeted victim $i$ has the maximum loss. 
We therefore ignore the maximum in the most part of the remaining paper.

\begin{lemma}
    \label{lem:Ulv}
$U(p_r ,S^l_{j \rightarrow i})= U(p_r ,S^v_{i \rightarrow j})$ holds for any player $p_r$ including $p_i$ and $p_j$.
\end{lemma}
This Lemma follows from the fact that, if $p_i$'s signature is missing in the block, rewards are the same, disrespect of wether the signature was omitted by the leader $p_j$ or delayed by the validator $p_i$.
It can be proven by considering Equation~(\ref{eqU:lj}) and~(\ref{eqU:li}) and setting up similar equations for $S^v_{i\rightarrow j}$.

\begin{theorem}
    \label{thm:costbalance}
The cost of vote omission and vote delay are inverse
    \[\cost(S^l_{j\rightarrow i})=\frac{1}{\cost(S^v_{i\rightarrow j})}\]
\end{theorem}

Theorem~\ref{thm:costbalance} shows that by designing a smart rewarding function $R$, a protocol designer can at most hope to balance the \cost of the vote omission and vote delay attacks. 
Making one attack unattractive by forcing the attacker to take a large cost, will also make the other attack appealing. 
A balanced \cost of 1 means that the attacker looses the same amount as its victim.
Accordingly, we define that:
\begin{definition}
We say that the attack cost is \emph{balanced}, if $$\cost(S^l_{j\rightarrow i})=\cost(S^v_{j\rightarrow i})=1$$
\end{definition}

\begin{theorem}
    \label{thm:effbalance}
The \effectiveness of vote omission and vote delay is related as follows:
\[
\frac{\effectiveness(S^l_{j\rightarrow i})}{\effectiveness(S^v_{i\rightarrow j})}= \cost(S^v_{i\rightarrow j})\cdot \frac{U(p_j,S_e)\pow[i]}{U(p_i,S_e)\pow[j]}
\]
\end{theorem}

Theorem~\ref{thm:effbalance} shows that also the effectiveness of the two attacks is related.
Note that, if the utility different players receive in $S_e$ is directly related to their power, then the last term in Theorem~\ref{thm:effbalance} becomes 1.
The Theorem follows when inserting Lemma~\ref{lem:Ulv} into the definition of \effectiveness.

\begin{example}
    \label{exampleCost}
Given the reward function $R$ from Example~\ref{exampleR}, a balanced attack cost of is achieved by setting $b=1$. 
Thus the reward for the leader has to be as big as the total reward for all validators.

\end{example}

\subsection{Protection Mechanisms}
\label{sec:protection}
In this section we review multiple mechanisms that different systems use to discourage vote omission.
The mechanisms allow to discourage vote omission even with a smaller bonus.

As we showed in Example~\ref{exampleCost}, using a simple rewarding function, a particularly large bonus is needed to give vote omission a cost of 1 or more.
In practice, such a large leader bonus may be considered problematic. 
It does not reflect the additional work required from the leader. 
Further, such a large leader bonus may provoke other attacks on the leader. 
Finally, since significant time may pass, between two times that a validator becomes the leader, such a large leader reward introduces high variability in a validators reward. 

We discuss individual mechanisms and if and how the mechanisms can be modelled as update to function $R(\delta_l,\delta_i, \pow[i], \powersum)$ from Example~\ref{exampleR}.
We also discuss how the mechanisms affect the balance between \cost and \effectiveness of the two attacks, shown in Theorem~\ref{thm:costbalance} and~\ref{thm:effbalance}.
In the following sections we analyse the specific combination and parametrization of mechanisms in different systems.
Table~\ref{table:mechanisms} summarizes the mechanisms and their effects.

\begin{itemize}
    \item{\textbf{\threshold}}
    This mechanism introduces a threshold designating a maximum fraction of signatures that a leader must collect before receiving any bonus. 
    The reasoning behind this mechanism is that many algorithms require at last 2/3 of the validators' signatures to commit a block and include it in the blockchain.
    Thus, a block with signatures from less than 2/3 of the validators may not even be included. 
    Additionally, to incentivize leaders to collect more than the minimum amount of signatures the bonus is scaled based purely on the additional signatures found.

    Let $t$ describe the bonus threshold. Thus, if fewer than $t$ signatures are collected ($\powersum < t$), all players receive a zero reward. 
    If more signatures are collected, the leaders bonus is scaled based on the additional included signatures.
    Thus, the reward function from Example~\ref{exampleR} is adjusted to:
    \begin{equation}
        \label{eq:R:t}
        R(\delta_l,\delta_i, \pow[i], \powersum)=
            \delta_l \cdot \frac{\powersum-t}{1-t} \cdot bR + \delta_i \cdot \pow[i] \cdot R
    \end{equation}

    The bonus threshold increases the cost of vote omission by the leader.
    For example with a threshold $t=2/3$, when omitting votes with power $\pow[i]=0.1$, the leader looses 30\% of its reward.
    Using the reward function from Equation~(\ref{eq:R:t}), the attack cost and efficiency is balanced, if $b=1-t$. 

    The Cosmos~\cite{cosmos2024validator} blockchain uses a threshold $t=2/3$.

    \item \textbf{\Scaling}: 
    In \scaling, the reward any player receives is scaled with the fraction of votes included in the block.
    Thus, missing signatures do not only affect the leaders bonus, but all validators rewards.
    Especially, with \scaling, attacks also reduce the utility received by other players.
    %
    \Scaling does not use additional parameters. 
    The scaled reward $R$ can be computed based on the reward without scaling, as shown below. Note that here, since the leaders bonus is scaled from before, no additional scaling is applied.
    \begin{equation}
        R(\delta_l,\delta_i, \pow[i], \powersum)=
        \delta_l \cdot \powersum \cdot bR + \powersum \cdot \delta_i \cdot \pow[i] \cdot R
    \end{equation}

    \Scaling does not change the \effectiveness of vote omission, since the victims reward anyway is zero in rounds its signatures are omitted.
    However, it does increase the \cost of the vote omission attack since the attacker also looses part of the reward, received for including his own signature. 
    Thus, an attacker with larger power has a bigger cost than a small attacker, to do the same harm.
    The cost for vote omission by player $p_j$ is given by $b+\pow[j]$.
    According to Theorem~\ref{thm:costbalance}, the \cost of the vote delay attack is decreased. 
    Also, the attack is more effective against victims with large power.
    Since the cost depends on the attackers or victims power, with \scaling, it is not possible to ensure that the attack cost is balanced for attackers of different power.

    \Scaling is used in Ethereum~\cite{ethereum2024rewards}.

    \item \textbf{\Window}: 
    \Window is a mechanism that allows votes for one block to be included in a successive block. 
    If a vote is included in a later block, the reward is reduced, but not lost completely. 
    This reduces the \effectiveness of the vote omission attack, since the omitted vote will be included later.
    It also reduces the cost of the vote delay attack, since the attacker can get his signature included in a later block.

    \Window is parametrized by the reduction factor $\rho$ and the window size $w$. 
    The window size $w$ determines that a vote can be included in the next $w$ blocks. 
    The window size should be chosen large enough, to prevent that an attacker controls $w$ consecutive leaders. 
    If a vote is included late, the validator receives the original reward, multiplied by~$\rho$.

    If we assume that $w$ is large enough, such that a delayed or omitted vote always is included in a later block, by a leader controlled by a different player, we can adjust the reward function from Example~\ref{exampleR} as follows:
    \begin{equation}
        \label{eq:R:w}
        R(\delta_l,\delta_i, \pow[i], \powersum)=
            \delta_l \cdot \powersum \cdot bR + \rho \pow[i] \cdot R  + \delta_i (1-\rho) \cdot \pow[i] \cdot R 
    \end{equation}
    Note that, even if a vote of a non-leader is not included, the player receives $\rho \pow[i] \cdot R$.

    \Window is used in Ethereum~\cite{ethereum2024rewards} with $w=6$ and $\rho=0.781$.
    With $w=6$, the probability that an attacker with $\pow[i]=1/3$ can omit a victim from $w$ consecutive blocks is less than 0.002.

    \item \textbf{\Basereward}: 
    \Basereward is a mechanism that ensures that validators receive a fraction of the rewards even if their vote is not included.
    This reduces the \effectiveness of the vote omission attack, similar to \window.
    Compared to \window, \basereward does create less overhead for the protocol, since it does not require to detect and reward late votes. 
    However, \basereward does reduce the motivation for validators to run highly reliable hardware.

    \Basereward uses a parameter $a$ that indicates the fraction of the reward a validator receives, even if his vote is not included.

    \Basereward reduces the \effectiveness of the vote omission attack to $1-a$. 

    A variant of \basereward is used in Cosmos~\cite{kwon2016cosmos}, where the block reward is distributed to all eligible validators for the block, while the transaction fees are only given to those who have voted for the block.
    In Cosmos, transaction fees typically only make up $10\%$ of a validators reward.

    \item \textbf{\Aggregation}:
    \Aggregation separates responsibilities between aggregators, that collect votes, and the leader, that includes the votes in the block. 
    Votes collected by the aggregators are combined in a way that the leader cannot omit individual ones.
    There may be multiple aggregators, in which case the leader can decide to include one or more of the aggregates in the block.
    With \Aggregation, an attacker may not be able to omit votes whenever he controls the leader.
    The attacker may only omit votes if he controls both the leader and one aggregator. 
    Additionally, if the attacker controls all aggregators, he may omit votes, even if he does not control the leader.

    \Aggregation is parametrized by $k$, the number of aggregators.
    A small $k$ makes it more likely that the attacker can control all aggregators. 
    A large $k$ makes it more likely, that the attacker controls at least one aggregator, in cases he controls the leader.
    
    \Aggregation reduces \effectiveness, but does not impact cost. 
    Different from the other mechanisms presented here, \Aggregation does not change the reward function $R$ but instead makes it harder to at all perform vote omission.
    Therefore this mechanisms avoids the balance in Theorems~\ref{thm:costbalance} and~\ref{thm:effbalance}. It reduces the \effectiveness of vote omission, without making vote delay more attractive.
    If we assume that the probability for one aggregator, to be controlled by the attacker is equal to the attackers power $\pow[a]$, and that this is independent for multiple aggregators, the probability for the attacker to perform vote omission becomes $\pow[a](1-(1-\pow[a])^k)+\pow[a]^{k}$. 
    Assuming independence is justified if the number of validators the attacker controls is significantly larger than the number of aggregators.
    The \effectiveness is reduced most, for $k=2$ or $k=3$, but the effect is very different depending on the attackers power.

    \Aggregation is used in Ethereum~\cite{edgington2023aggregator}.
    The number of aggregators in Ethereum may vary but on average $k=16$ is used.
    This has little effect on large attackers, but reduces the \effectiveness for small attackers.

\end{itemize}

\begin{table}[t]
\centering
\renewcommand{\arraystretch}{2} 
\begin{tabular}{|c|c|c|c|c|}
\hline
\multirow{2}{*}{\textbf{Attack Type}} & \multicolumn{2}{c|}{\textbf{Vote delay}} & \multicolumn{2}{c|}{\textbf{Vote omission}} \\ \cline{2-5} 
                                      & \textbf{\Cost}      & \textbf{\Effectiveness}    & \textbf{\Cost}      & \textbf{\Effectiveness}     \\ \hline
\textbf{\Threshold}                      & 
\arrowdown \hspace{0.5em} \Sadey[3]       & 
\arrowup \hspace{0.5em} \Sadey[3]       & 
\arrowup \hspace{0.5em} \Smiley[3]     & 
\arrowdown \hspace{0.5em} \Smiley[3]     \\ \hline

\textbf{\Scaling}                      & 
{-}       & 
\arrowup \hspace{0.5em} \Sadey[3]       & 
\arrowup \hspace{0.5em} \Smiley[3]     & 
{-}     \\ \hline

\textbf{\Window}                      & 
\arrowdown \hspace{0.5em} \Sadey[3]       & 
-       & 
-    & 
\arrowdown \hspace{0.5em} \Smiley[3]      \\ \hline

\textbf{\Aggregation}                      & 
-       & 
-       & 
-    & 
\arrowdown \hspace{0.5em} \Smiley[3]      \\ \hline

\textbf{\Basereward}             & 
\arrowdown \hspace{0.5em} \Sadey[3]       & 
-       & 
-     & 
\arrowdown \hspace{0.5em} \Smiley[3]     \\ \hline
\end{tabular}
\caption{Different mechanisms' effect on \Cost and \Effectiveness of Vote Delay vs. Vote Omission Attacks. 
A reduction in \Cost makes the attack cheaper and the more likely, while a reduction in \Effectiveness makes the attack less effective.}
\label{table:mechanisms}
\end{table}


Table~\ref{table:mechanisms} summarises the above mechanisms. 
\Threshold and \scaling increase the \cost of the attack.
\window and \basereward allow the victim to receive a fraction of his reward, even if its vote is omitted. They thus reduce the \effectiveness of the attack. 
Since these mechanisms only change the rewarding function, they are subject to Theorem~\ref{thm:costbalance}, and by making one attack less attractive they make another attack more attractive. 
However, these mechanisms can be used to achieve balance with a somewhat smaller bonus.
Different from the other mechanisms, \aggregation avoids the balance.
\Aggregation makes it more difficult for the attacker to omit votes, and thus also reduces effectiveness, without affecting vote delay.

\section{Systems Analysis}
\label{sec:usecases}
In this section, we use our framework to analyze the validator rewards in two well-known committee-based blockchains, Cosmos~\cite{kwon2016cosmos} and Ethereum~\cite{buterin2018ethereum}. 

\subsection{Cosmos Rewarding Analysis}
\label{sec:cosmos}
In the following we model and analyze the validator rewards in Cosmos. 
For the parameters used in Cosmos, our analysis shows a strong imbalance between the vote omission and vote delay attacks, making vote omission not only powerful but even profitable.  
We propose alternative parameters that make the scheme an equilibrium and analyze the resulting balance.

In Cosmos, rewards consist of two main parts: transaction fees paid by users and a block reward of newly minted tokens.
While the block reward stays constant during short periods of time and solely depends on the inflation rate and total Atom supply, transaction fees vary and depend on the transactions included in the block. 
The block reward is distributed among all validators, regardless of whether they have voted or not~\cite{cosmos-sdk-distribution-beginblock-v046}. 
This corresponds to the \basereward introduced in Section~\ref{sec:vote}.

The transaction fees are distributed only among the validators whose signatures are included. 
Therefore, if a signature is missing from a block, the corresponding validator loses its share of the transaction fees.
Cosmos redistributes the lost shares among all validators instead of burning them. 
A portion of the transaction fees (5\%) is allocated to the proposer as a bonus. 
However, the bonus is also reduced if some signatures are missing and the resulting funds are similarly redistributed among all validators.
Cosmos uses a \textit{bonus threshold} of $2/3$. 

\begin{figure}
    \centering
    \subfloat[Rewarding scheme in Cosmos without \dop attacks. Proposer receives bonus only for votes above the threshold.]{        \begin{tikzpicture}[node distance=0cm,scale=0.9]

            \colorlet{color1}{cyan}
            \colorlet{color2}{green}
            \colorlet{color3}{yellow!70!gray}
            \colorlet{color4}{orange}
            \colorlet{color5}{magenta}
            \colorlet{colorx}{gray}
            
            \newcommand{\emojihead}[3]{%
              \node at (#1, -2) {\textcolor{#2}{\scalebox{0.8}{#3}}};
            }
            \newcommand{\cancel}[2]{%
              \node[draw, cross out, ultra thick] at (#1, #2) {};
            }
            
            \emojihead{0}{color1}{\Smiley[3]}
            \emojihead{1}{color2}{\Smiley[3]}
            \emojihead{2}{color3}{\Smiley[3]}
            \emojihead{3}{color4}{\Smiley[3]}
            \emojihead{4}{color5}{\Smiley[3]}
            
            \node (prop) at (-1, 1) {\scalebox{0.8}{\textcolor{black}{\Smiley[3]}}};
            \node[single arrow, draw, minimum height=4.4cm, minimum width=1cm,shape border rotate=180,
                single arrow head extend=0.15cm,
            ] (arr) [right=of prop] {};
            
            \newcommand{\letter}[4][0]{%
              \node at (#2, #1) {\textcolor{#3}{\huge $\mathcal{#4}$}};
            };

            \newcommand{\letters}[4]{%
              \node at (#1, #2) {\textcolor{#3}{$\mathcal{#4}$}};
            };

            \letter{0}{color1}{A};
            \letter{1}{color2}{B};
            \letter{2}{color3}{C};
            \letter{3}{color4}{D};
            \letter{4}{color5}{E};




            \draw[ultra thick] (-0.5, -0.5) rectangle (4.5, 0.5);
            \draw[ultra thick] (-1.5, -0.5) rectangle (-1, 0.5);
            \draw[ultra thick] (-1, 0) -- (-0.5, 0);
            \draw[color=white,fill=white] (-1.6, -0.6) rectangle (-1.4, 0.6);
            

            \newcommand{\coloredcircle}[4][1]{%
                \node[draw, circle, fill=#2, minimum size=0.2cm, inner sep=0.02cm] at (#3, #4) {\scalebox{#1}{\textcolor{black}{\textdollar}}};
            }

            \newcommand{\coloredcirclesmall}[4][0.7]{%
                \node[draw, circle, fill=#2, minimum size=0.1cm, inner sep=0.01cm] at (#3, #4) {\scalebox{#1}{\textcolor{black}{\textdollar}}};
            }

            \newcommand{\circleslice}[3]{%
                \node[draw, circular sector, fill=#2,
                minimum size=0.12cm,
                circular sector angle=72,
                rotate=54,
                inner sep=0.01cm] at (#1, #3) {};
            }

            \newcommand{\circleslicesmall}[3]{%
                \node[draw, circular sector, fill=#2,
                minimum size=0.08cm,
                circular sector angle=72,
                rotate=54,
                inner sep=0.005cm] at (#1, #3) {};
            }

            \newcommand{\coloredcirclex}[4][1]{%
                \node[draw, circle, fill=#2, minimum size=0.2cm, inner sep=0.02cm] at (#3, #4) {\textcolor{black}{\textdollar}};
                \node[fill=white, circular sector,circular sector angle=72,rotate=-54, inner sep=0.07cm] (cp) [left={0cm of (#3, #4)}] {};
            }
            \newcommand{\coloredcircley}[4][1]{%
                \node[draw, circle, fill=#2, minimum size=0.2cm, inner sep=0.02cm] at (#3, #4) {\scalebox{0.6}{\textcolor{black}{\textdollar}}};
                \node[fill=white, circular sector,circular sector angle=72,rotate=-54, inner sep=0.045cm] (cp) [left={0cm of (#3, #4)}] {};
            }
            \newcommand{\coloredcirclez}[4][1]{%
                \node[draw, circle, fill=#2, minimum size=0.2cm, inner sep=0.02cm] at (#3, #4) {\textcolor{black}{\textdollar}};
                \node[fill=white, circular sector,circular sector angle=90,rotate=45, inner sep=0.08cm] (cp) [left={0cm of (#3, #4)}] {};
            }

            \foreach \x/\col in {0/color1, 1/color2, 2/color3, 3/color4, 4/color5} {
                \draw[->, thick] (\x, -0.3) -- (\x, -1.5);
                \coloredcircle{\col}{\x - 0.2}{-1};          
            };

            \foreach \x/\col in {0/color1, 1/color2, 2/color3, 3/color4, 4/color5} {
                \coloredcirclesmall{\col}{\x - 0.60}{-1};    
            };


            \foreach \x/\col in {3/color4} {
                \coloredcircle[0.4]{\col}{\x}{1};
            };
            \coloredcircle[0.4]{color5}{4}{1};
            \node at (2,-2.7) {validators};
            \node [above=-0.2cm of prop] {proposer};
            
        \end{tikzpicture}}
    \hfill
    \subfloat[\dop attacks in Cosmos: Vote $\mathcal{E}$ is omitted or delayed and reward and bonus for this vote are redistributed.
    Validator $\mathcal{E}$ still receives the \basereward.]{        \begin{tikzpicture}[node distance=0cm,scale=0.9]

            \colorlet{color1}{cyan}
            \colorlet{color2}{green}
            \colorlet{color3}{yellow!70!gray}
            \colorlet{color4}{orange}
            \colorlet{color5}{magenta}
            \colorlet{colorx}{gray}
            
            \newcommand{\emojihead}[3]{%
              \node at (#1, -2) {\textcolor{#2}{\scalebox{0.8}{#3}}};
            }
            \newcommand{\cancelbig}[2]{%
              \node[draw, cross out, ultra thick] at (#1, #2) {};
            }
            \newcommand{\cancel}[2]{%
              \node[draw, cross out, very thick] at (#1, #2) {};
            }
            
            \emojihead{0}{color1}{\Laughey[3]}
            \emojihead{1}{color2}{\Laughey[3]}
            \emojihead{2}{color3}{\Laughey[3]}
            \emojihead{3}{color4}{\Laughey[3]}
            \emojihead{4}{color5}{\Sey[3]}
            
            \node (prop) at (-1, 1) {\scalebox{0.8}{\textcolor{black}{\Sey[3]}}};
            \node[single arrow, draw, minimum height=4.4cm, minimum width=1cm,shape border rotate=180,
                single arrow head extend=0.15cm,
            ] (arr) [right=of prop] {};
            
            \newcommand{\letter}[4][0]{%
              \node at (#2, #1) {\textcolor{#3}{\huge $\mathcal{#4}$}};
            };

            \newcommand{\letters}[4]{%
              \node at (#1, #2) {\textcolor{#3}{$\mathcal{#4}$}};
            };

            \letter{0}{color1}{A};
            \letter{1}{color2}{B};
            \letter{2}{color3}{C};
            \letter{3}{color4}{D};
            \letter{4}{colorx}{E};
            \cancelbig{4}{0};




            \draw[ultra thick] (-0.5, -0.5) rectangle (4.5, 0.5);
            \draw[ultra thick] (-1.5, -0.5) rectangle (-1, 0.5);
            \draw[ultra thick] (-1, 0) -- (-0.5, 0);
            \draw[color=white,fill=white] (-1.6, -0.6) rectangle (-1.4, 0.6);
            

            \newcommand{\coloredcircle}[4][1]{%
                \node[draw, circle, fill=#2, minimum size=0.2cm, inner sep=0.02cm] at (#3, #4) {\scalebox{#1}{\textcolor{black}{\textdollar}}};
            }

            \newcommand{\coloredcirclesmall}[4][0.7]{%
                \node[draw, circle, fill=#2, minimum size=0.1cm, inner sep=0.01cm] at (#3, #4) {\scalebox{#1}{\textcolor{black}{\textdollar}}};
            }

            \newcommand{\circleslice}[3]{%
                \node[draw, circular sector, fill=#2,
                minimum size=0.12cm,
                circular sector angle=72,
                rotate=54,
                inner sep=0.01cm] at (#1, #3) {};
            }

            \newcommand{\circleslicesmall}[3]{%
                \node[draw, circular sector, fill=#2,
                minimum size=0.08cm,
                circular sector angle=72,
                rotate=54,
                inner sep=0.005cm] at (#1, #3) {};
            }

            \newcommand{\coloredcirclex}[4][1]{%
                \node[draw, circle, fill=#2, minimum size=0.2cm, inner sep=0.02cm] at (#3, #4) {\textcolor{black}{\textdollar}};
                \node[fill=white, circular sector,circular sector angle=72,rotate=-54, inner sep=0.07cm] (cp) [left={0cm of (#3, #4)}] {};
            }
            \newcommand{\coloredcircley}[4][1]{%
                \node[draw, circle, fill=#2, minimum size=0.2cm, inner sep=0.02cm] at (#3, #4) {\scalebox{0.6}{\textcolor{black}{\textdollar}}};
                \node[fill=white, circular sector,circular sector angle=72,rotate=-54, inner sep=0.045cm] (cp) [left={0cm of (#3, #4)}] {};
            }
            \newcommand{\coloredcirclez}[4][1]{%
                \node[draw, circle, fill=#2, minimum size=0.2cm, inner sep=0.02cm] at (#3, #4) {\textcolor{black}{\textdollar}};
                \node[fill=white, circular sector,circular sector angle=90,rotate=45, inner sep=0.08cm] (cp) [left={0cm of (#3, #4)}] {};
            }

            \foreach \x/\col in {0/color1, 1/color2, 2/color3, 3/color4, 4/color5} {
                \draw[->, thick] (\x, -0.3) -- (\x, -1.5);
                \coloredcircle{\col}{\x - 0.2}{-1};          
                \circleslice{\x - 0.2}{\col}{-1.4};  
                \circleslicesmall{\x - 0.40}{\col}{-1.4}; 
            };

            \foreach \x/\col in {0/color1, 1/color2, 2/color3, 3/color4} {
                \coloredcirclesmall{\col}{\x - 0.60}{-1};    
            };
            \coloredcirclesmall{white}{3.4}{-1}; 

            \node[draw=color1, circular sector, fill=color1,
                minimum size=0.11cm,
                circular sector angle=72,
                rotate=54,
                inner sep=0.005cm,anchor=sector center] at (3.4,-1) {};
            \node[draw=color2, circular sector, fill=color2,
                minimum size=0.11cm,
                circular sector angle=72,
                rotate=-18,
                inner sep=0.005cm,anchor=sector center] at (3.4,-1) {};

            \node[draw=color3, circular sector, fill=color3,
                minimum size=0.11cm,
                circular sector angle=72,
                rotate=-90,
                inner sep=0.005cm,anchor=sector center] at (3.4,-1) {};
            \node[draw=color4, circular sector, fill=color4,
                minimum size=0.11cm,
                circular sector angle=72,
                rotate=-162,
                inner sep=0.005cm,anchor=sector center] at (3.4,-1) {};
            \node[draw=color5, circular sector, fill=color5,
                minimum size=0.11cm,
                circular sector angle=72,
                rotate=126,
                inner sep=0.005cm,anchor=sector center] at (3.4,-1) {};

            \node[draw, circle, minimum size=0.1cm, inner sep=0.01cm] at (3.4, -1) {\scalebox{0.7}{\textcolor{black}{\textdollar}}};
            \cancel{3.4}{-1};


            \foreach \x/\col in {3/color4} {
                \coloredcircle[0.4]{\col}{\x}{1};
            };
            \coloredcircle[0.4]{color5}{4}{1};
            \node[draw=color1, circular sector, fill=color1,
                minimum size=0.08cm,
                circular sector angle=72,
                rotate=54,
                inner sep=0.005cm,anchor=sector center] at (4,1) {};
            \node[draw=color2, circular sector, fill=color2,
                minimum size=0.08cm,
                circular sector angle=72,
                rotate=-18,
                inner sep=0.005cm,anchor=sector center] at (4,1) {};

            \node[draw=color3, circular sector, fill=color3,
                minimum size=0.08cm,
                circular sector angle=72,
                rotate=-90,
                inner sep=0.005cm,anchor=sector center] at (4,1) {};
            \node[draw=color4, circular sector, fill=color4,
                minimum size=0.08cm,
                circular sector angle=72,
                rotate=-162,
                inner sep=0.005cm,anchor=sector center] at (4,1) {};
            \node[draw=color5, circular sector, fill=color5,
                minimum size=0.08cm,
                circular sector angle=72,
                rotate=126,
                inner sep=0.005cm,anchor=sector center] at (4,1) {};
            \node[draw, circle, minimum size=0.1cm, inner sep=0.01cm] at (4, 1) {\scalebox{0.5}{\textcolor{black}{\textdollar}}};

            \cancel{4}{1};
            \node at (2,-2.7) {validators};
            \node [above=-0.2cm of prop] {proposer};
            
        \end{tikzpicture}}
    \caption{Cosmos rewarding scheme with and without attack.} 
    \label{fig:cosmos-rewarding}
\end{figure}
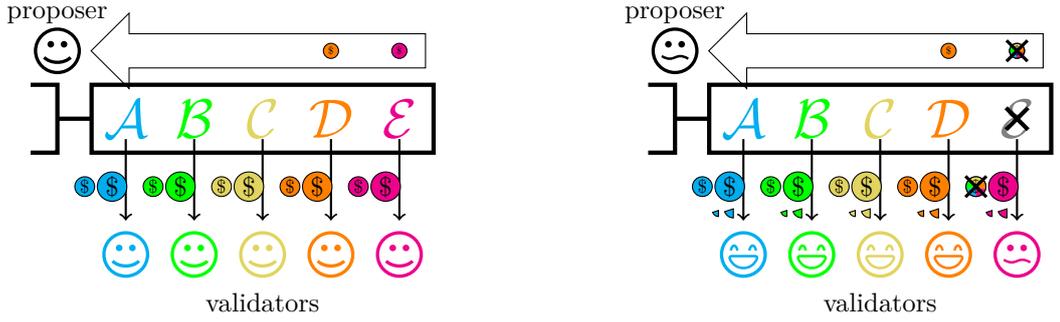

We model the Cosmos reward scheme as a vote collection game as described in Section~\ref{sec:game}, and analyze the attacks discussed. 
Figure~\ref{fig:cosmos-rewarding} illustrates the rewarding scheme and the redistribution of lost transaction fee shares.
To simplify the analysis, we assume that transaction fees are constant.
We use the parameters $a$, $b$, and $t$ to denote the \basereward fraction, the proposer bonus fraction, and the bonus threshold, in accordance with the mechanisms introduced in Section~\ref{sec:protection}.
Additionally, we model the redistribution in Cosmos.

\subsubsection{Reward function}

Utilities in Cosmos follow the Equations~(\ref{eqU:lj}) and (\ref{eqU:li}).
Thus, it is sufficient to define the rewarding function $R(\delta_l,\delta_i, \pow[i], \powersum)$.
For every validator $p_i$ in Cosmos, the reward function $R$ is derived as follows:

\begin{align}
    \label{eq:cosmosR}
    R(\delta_l,\delta_i, \pow[i], \powersum) =\ &     
    \delta_l \cdot \frac{\powersum - t}{1-t} \cdot b(1-a)R  \nonumber \\
    &+ \delta_i \cdot \pow[i] \cdot (1-a)(1-b)R \nonumber \\
    &+ \pow[i] \cdot aR \nonumber \\
    &+ \pow[i] \cdot \left( 1-\frac{\powersum - t}{1-t}\right) \cdot b(1-a)R \nonumber \\
    &+ \pow[i] \cdot (1 - \powersum) \cdot (1-a)(1-b)R
\end{align}

Here, the first three summands express the leader's bonus, the voting reward, and the base reward.
The last two summands are the result of redistributed bonus and voting reward, in case some signatures are omitted.

By inserting victim and attacker utilities derived by $R$ into respective definitions, \effectiveness and \cost of vote omission attack are derived as follows: 
\begin{equation}
    \label{eq:cosmos-vo-eff}
    \effectiveness(S^l_{i\rightarrow j}) = 
\frac{(1-a)}{(1-t)}
\Bigl[
  (1-t)\,(1-b)\,(1 - \pow[j])
  -
  \pow[j]\,b
\Bigr]
\end{equation}

\begin{equation}
    \label{eq:cosmos-vo-cost}
    \cost(S^l_{i\rightarrow j}) = 
\frac{b(1-\pow[i])- \pow[i]\,(1-t)\,(1-b)}{(1-t)(1-b)(1-\pow[j])- \pow[j]\,b}\,
\end{equation}
\effectiveness and \cost of vote delay attack can be derived from these Equations, according to Theorem~\ref{thm:effbalance} and Theorem~\ref{thm:costbalance}.

\subsubsection{Attacks in Cosmos}

\begin{figure}[ht]
\includegraphics[width=\linewidth]{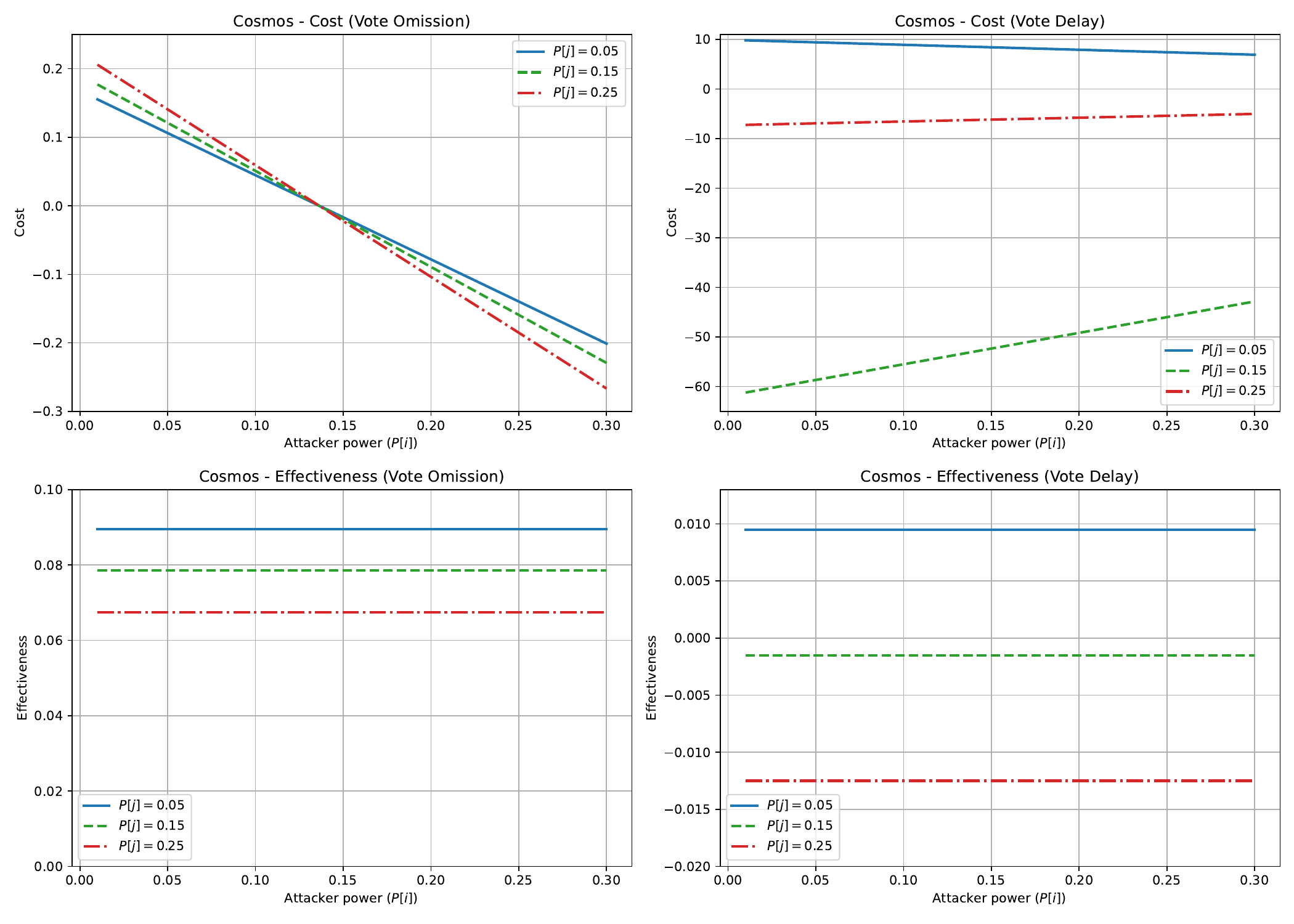}
\caption{
\textbf{Analysis of Cosmos Attack Metrics versus Attacker Power.}
 The left panel shows the \cost of vote omission (solid lines) and vote delay (dotted lines) attacks as a function of attacker power ($\pow[i]$) for three victim stake values ($\pow[j] = 0.05, 0.15, and 0.25$). The right panel displays the corresponding \effectiveness for both types of attacks, which remain invariant with respect to attacker power.
 Parameters: \( t = \frac{2}{3} \), \( a = 0.9 \), \( b = 0.05 \).
}
\label{fig:cosmos-analysis-m}
\end{figure}

Figure~\ref{fig:cosmos-analysis-m} illustrates how \cost and \effectiveness of vote omission and vote delay attacks in Cosmos vary based on the attacker power ($\pow[i]$) and victim power($\pow[j]$).
For vote omission (strategy $S_{i\rightarrow j}^l)$, \cost decreases if $\pow[i]$ increases. 
This shows that a powerful attacker can perform the attack cheaper. 
In addition, attacking a smaller victim increases both the relative damage (\effectiveness) and reduces the \cost for the attacker, making it much cheaper and more lucrative to attack small victims.  
However, in vote delay (strategy $S_{i\rightarrow j}^v)$, the \effectiveness gets negative if $\pow[j]$ increases.
This means the victim actually gains from the attack in these cases. 
The negative \cost in that case shows the attacker loss per victim gain. 

According to Figure~\ref{fig:cosmos-analysis-m}, given the parameters used in Cosmos, attackers can benefit from omitting votes in this design.
This means that following this protocol is not a Nash Equilibrium under these parameters, as shown in previous work~\cite{lagaillardie2019computational,rebop,eiffel}. 
The cost of vote omission is negative for attackers which control more than  14\% of the stake, indicating that this attack is profitable for these attackers.
Further, omitting votes from larger victims gives a larger gain.
On the other hand, the vote delay attack never becomes profitable.
This imbalances in \cost and \effectiveness of the two attacks leads to vulnerabilities in the model.
In 2024, Cosmos revised its reward mechanism and removed the proposer bonus entirely, setting it to $0$~\cite{cosmos-sdk-changelog, cosmos-sdk-distribution-v050}.\footnote{We discuss the earlier version, since it has better documentation available.}
While this change simplifies the fee distribution logic, it also eliminates the incentive for proposers to maximize inclusion of validator signatures, making omission attacks more attractive, rather than less.
Thus, the change does not directly mitigate the vulnerabilities described here, and may, in fact, amplify them.

\subsubsection{Alternative parameters}

Note that \cost and \effectiveness of the attacks depend on several parameters, including the bonus value $b$. 
For the vote omission attack, as $b$ increases, \effectiveness decreases, while \cost increases. 
It is the opposite for vote delay attack. 

\begin{corollary}
\label{cor:cos-balanced-b}
Numerical analysis shows that in the Cosmos model with parameters $a=0.9$ and $t=\tfrac23$, the minimum bonus value $b^ * \approx 0.141$ guarantees that for all attacker and victim powers $\pow[i],\pow[j]$ the \cost for both vote omission and vote delay attacks is larger than $0$, so the correct strategy profile is a Nash equilibrium.
\end{corollary}

Corollary~\ref{cor:cos-balanced-b} shows that the minimum proposer bonus $b^*\approx0.141$ needed to make both vote omission and vote delay unprofitable is far above the $b=0.05$ currently chosen by Cosmos.
This suggests that the current reward scheme is unbalanced.
One possible explanation is that the designers deliberately kept $b$ low because they believe a smaller proposer bonus better reflects the actual work and responsibility of block creation.
However, we note that due to the base reward, which is given unconditionally, even with the larger $b^*$, the bonus actually only makes up $1.4\%$ of the total reward paid out for the block.

Note that while $b^ *$ guarantees that the correct strategy profile is a Nash equilibrium, a larger bonus is needed to balance the attacks. 
For example, a bonus of $0.21$ ensures that \cost for both attacks is larger than $0.5$ for all attackers and victims with power less than $1/3$.

\subsection{Ethereum Rewarding Analysis}
\label{sec:ethereum}
In 2022 Ethereum transitioned from a proof-of-work consensus algorithm to a proof-of-stake committee-based system~\cite{themerge}.
The new structure gives out rewards to validators based on inclusion in a block.
Ethereum uses \scaling, \aggregation, and \window mechanisms to mitigate the vote omission attack.

In Ethereum, not every validator votes for every block. 
Instead validators are randomly divided into 32 committees, each voting for a different block. 
Each committee also has its own leader, selected at random.
The 32 blocks are called an epoch.
In the next epoch, the committees are shuffled, and the process repeats.
Unless they have been slashed for misbehavior, all validators have the same stake, namely 32 ETH. 
Thus, when modeling Ethereum as a vote collection game, we can assume the power represents the fraction of validators controlled by a player. 
If the power of players we consider is large enough, we can assume that they control approximately the same fraction of validators in one committee as in the whole system.
Similarly, when analyzing attacks between players $p_i$ and $p_j$ we can ignore committees where not both players are present.
Further, the probability that player $p_i$ controls the leader in a committee where player $p_j$ controls a validator is still equal to $\pow[i]$.
Due to this, we assume a single committee where each player holds a constant amount of power.

\begin{figure}
    \centering
    \subfloat[Rewarding scheme in Ethereum without \dop attacks.]{        \begin{tikzpicture}[node distance=0cm,scale=0.9]

            \colorlet{color1}{cyan}
            \colorlet{color2}{green}
            \colorlet{color3}{yellow!70!gray}
            \colorlet{color4}{orange}
            \colorlet{color5}{magenta}
            \colorlet{colorx}{gray}
            
            \newcommand{\emojihead}[3]{%
              \node at (#1, -2) {\textcolor{#2}{\scalebox{0.8}{#3}}};
            }
            \newcommand{\cancel}[2]{%
              \node[draw, cross out, ultra thick] at (#1, #2) {};
            }
            
            \emojihead{0}{color1}{\Smiley[3]}
            \emojihead{1}{color2}{\Smiley[3]}
            \emojihead{2}{color3}{\Smiley[3]}
            \emojihead{3}{color4}{\Smiley[3]}
            \emojihead{4}{color5}{\Smiley[3]}
            
            \node (prop) at (-1, 1) {\scalebox{0.8}{\textcolor{black}{\Smiley[3]}}};
            \node[single arrow, draw, minimum height=4.4cm, minimum width=1cm,shape border rotate=180,
                single arrow head extend=0.15cm,
            ] (arr) [right=of prop] {};
            
            \newcommand{\letter}[4][0]{%
              \node at (#2, #1) {\textcolor{#3}{\huge $\mathcal{#4}$}};
            };

            \newcommand{\letters}[4]{%
              \node at (#1, #2) {\textcolor{#3}{$\mathcal{#4}$}};
            };

            \letter{0}{color1}{A};
            \letter{1}{color2}{B};
            \letter{2}{color3}{C};
            \letter{3}{color4}{D};
            \letter{4}{color5}{E};




            \draw[ultra thick] (-0.5, -0.5) rectangle (4.5, 0.5);
            \draw[ultra thick] (-1.5, -0.5) rectangle (-1, 0.5);
            \draw[ultra thick] (-1, 0) -- (-0.5, 0);
            \draw[color=white,fill=white] (-1.6, -0.6) rectangle (-1.4, 0.6);
            

            \newcommand{\coloredcircle}[4][1]{%
                \node[draw, circle, fill=#2, minimum size=0.2cm, inner sep=0.02cm] at (#3, #4) {\scalebox{#1}{\textcolor{black}{\textdollar}}};
            }

            \newcommand{\coloredcirclex}[4][1]{%
                \node[draw, circle, fill=#2, minimum size=0.2cm, inner sep=0.02cm] at (#3, #4) {\textcolor{black}{\textdollar}};
                \node[fill=white, circular sector,circular sector angle=72,rotate=-54, inner sep=0.07cm] (cp) [left={0cm of (#3, #4)}] {};
            }
            \newcommand{\coloredcircley}[4][1]{%
                \node[draw, circle, fill=#2, minimum size=0.2cm, inner sep=0.02cm] at (#3, #4) {\scalebox{0.6}{\textcolor{black}{\textdollar}}};
                \node[fill=white, circular sector,circular sector angle=72,rotate=-54, inner sep=0.045cm] (cp) [left={0cm of (#3, #4)}] {};
            }
            \newcommand{\coloredcirclez}[4][1]{%
                \node[draw, circle, fill=#2, minimum size=0.2cm, inner sep=0.02cm] at (#3, #4) {\textcolor{black}{\textdollar}};
                \node[fill=white, circular sector,circular sector angle=90,rotate=45, inner sep=0.08cm] (cp) [left={0cm of (#3, #4)}] {};
            }

            \foreach \x/\col in {0/color1, 1/color2, 2/color3, 3/color4, 4/color5} {
              \draw[->, thick] (\x, -0.3) -- (\x, -1.5);
              \coloredcircle{\col}{\x - 0.2}{-1};
            };


            \foreach \x/\col in {0/color1, 1/color2, 2/color3, 3/color4} {
                \coloredcircle[0.6]{\col}{\x}{1};
            };
            \coloredcircle[0.6]{color5}{4}{1};
            \node at (2,-2.7) {validators};
            \node [above=-0.2cm of prop] {proposer};
            
        \end{tikzpicture}}
    \hfill
    \subfloat[\dop attacks in Ethereum: Vote $\mathcal{E}$ is omitted or delayed and included in the next block. Validators' reward is scaled and bonus reduced.
    Validator $\mathcal{E}$ receives a fraction of the reward for late inclusion.]{        \begin{tikzpicture}[node distance=0cm,,scale=0.9]

            \colorlet{color1}{cyan}
            \colorlet{color2}{green}
            \colorlet{color3}{yellow!70!gray}
            \colorlet{color4}{orange}
            \colorlet{color5}{magenta}
            \colorlet{colorx}{gray}
            
            \newcommand{\emojihead}[3]{%
              \node at (#1, -2) {\textcolor{#2}{\scalebox{0.8}{#3}}};
            }
            \newcommand{\cancel}[2]{%
              \node[draw, cross out, ultra thick] at (#1, #2) {};
            }
            
            \emojihead{0}{color1}{\Neutrey[3]}
            \emojihead{1}{color2}{\Neutrey[3]}
            \emojihead{2}{color3}{\Neutrey[3]}
            \emojihead{3}{color4}{\Neutrey[3]}
            \emojihead{4}{color5}{\Sey[3]}
            
            \node (prop) at (-1, 1) {\scalebox{0.8}{\textcolor{black}{\Neutrey[3]}}};
            \node[single arrow, draw, minimum height=4.4cm, minimum width=1cm,shape border rotate=180,
                single arrow head extend=0.15cm,
            ] (arr) [right=of prop] {};
            
            \newcommand{\letter}[4][0]{%
              \node at (#2, #1) {\textcolor{#3}{\huge $\mathcal{#4}$}};
            };

            \newcommand{\letters}[4]{%
              \node at (#1, #2) {\textcolor{#3}{$\mathcal{#4}$}};
            };

            \letter{0}{color1}{A};
            \letter{1}{color2}{B};
            \letter{2}{color3}{C};
            \letter{3}{color4}{D};
            \letter{4}{colorx}{E};
            \cancel{4}{0};




            \draw[ultra thick] (-0.5, -0.5) rectangle (4.5, 0.5);
            \draw[ultra thick] (5, -0.5) rectangle (6.5, 0.5);
            \draw[ultra thick] (4.5, 0) -- (5, 0);
            \draw[color=white,fill=white] (6.4, -0.6) rectangle (6.6, 0.6);
            
            \node [right={of (5.7,0)}, align=center]{next\\ block};

            \newcommand{\coloredcircle}[4][1]{%
                \node[draw, circle, fill=#2, minimum size=0.2cm, inner sep=0.02cm] at (#3, #4) {\scalebox{#1}{\textcolor{black}{\textdollar}}};
            }

            \newcommand{\coloredcirclex}[4][1]{%
                \node[draw, circle, fill=#2, minimum size=0.2cm, inner sep=0.02cm] at (#3, #4) {\textcolor{black}{\textdollar}};
                \node[fill=white, circular sector,circular sector angle=72,rotate=-54, inner sep=0.07cm] (cp) [left={0cm of (#3, #4)}] {};
            }
            \newcommand{\coloredcircley}[4][1]{%
                \node[draw, circle, fill=#2, minimum size=0.2cm, inner sep=0.02cm] at (#3, #4) {\scalebox{0.6}{\textcolor{black}{\textdollar}}};
                \node[fill=white, circular sector,circular sector angle=72,rotate=-54, inner sep=0.045cm] (cp) [left={0cm of (#3, #4)}] {};
            }
            \newcommand{\coloredcirclez}[4][1]{%
                \node[draw, circle, fill=#2, minimum size=0.2cm, inner sep=0.02cm] at (#3, #4) {\textcolor{black}{\textdollar}};
                \node[fill=white, circular sector,circular sector angle=90,rotate=45, inner sep=0.08cm] (cp) [left={0cm of (#3, #4)}] {};
            }

            \foreach \x/\col in {0/color1, 1/color2, 2/color3, 3/color4} {
              \draw[->, thick] (\x, -0.3) -- (\x, -1.5);
              \coloredcirclex{\col}{\x - 0.2}{-1};
            };

            \letter{5.5}{color5}{E};
            \draw[->, thick] (5.5, -0.3) -- (4.3, -1.5);
            \coloredcirclez{color5}{4.5}{-1};

            \foreach \x/\col in {0/color1, 1/color2, 2/color3, 3/color4} {
                \coloredcircle[0.6]{\col}{\x}{1};
            };
            \coloredcircle[0.6]{color5}{4}{1};
            \cancel{4}{1};
            \node at (2,-2.7) {validators};
            \node [above=-0.2cm of prop] {proposer};
            
        \end{tikzpicture}}
    \caption{Ethereum rewarding scheme with and without attack.} 
    \label{fig:ethereum-rewarding}
\end{figure}
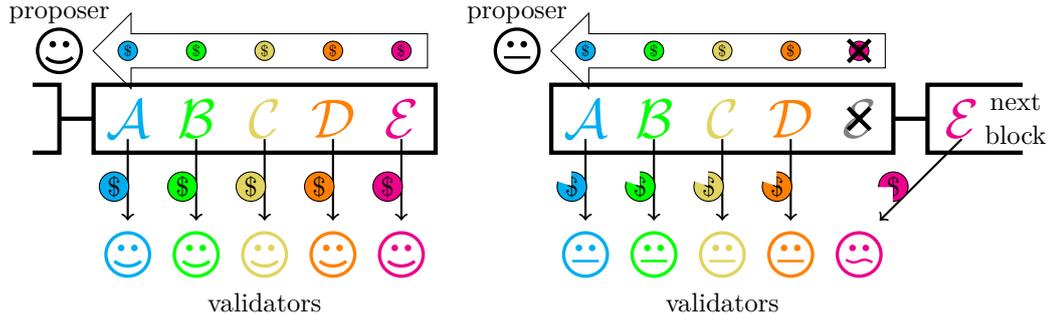

Figure \ref{fig:ethereum-rewarding} illustrates the rewarding mechanism in Ethereum.
We now discuss the protection mechanisms against vote omission attacks in Ethereum.

First, Ethereum uses \aggregation.
However, instead of directly aggregating one committee, Ethereum divides each committee into several sub-committees.
Within each sub-committee, some validators are selected as aggregators. 
These aggregators listen to the network and aggregate the votes from their respective sub-committees, then submit the aggregated votes to the network. 
As mentioned before, if the attacker controls either all aggregators in a subcommittee, or the leader and one aggregator, he can omit votes from that subcommittee.
Ethereum does not choose a fixed number of aggregators. 
Instead, every validator becomes an aggregator with a fixed probability, adjusted to give 16 aggregators on average. 
Thus, in a committee of size $c$ the probability $p_{agg}(a)$, that a player $p_a$ controls at least one aggregator is $1 - (1 - \frac{16}{c})^{P[a]c}$. Committee sizes in Ethereum can vary between $128$ and $2048$, but currently values lie around $c=500$. 
We note that for these values, the probability that an attacker controls all aggregators is minimal. We therefore ignore this case in our analysis.

Additionally, Ethereum employs the \window mechanism, where votes still receive a portion of the reward even if included in later blocks. 
As in Section~\ref{sec:protection} we assume that every omitted or delayed vote is included within the inclusion window.
We note that in theory, an attacker controlling two consecutive leaders could benefit from first omitting and later including a vote.
However, the chance to control leaders in consecutive blocks is small and we therefore ignore this case.

Ethereum uses \scaling to encourage all validators, including leader and aggregators to collect as many votes as possible. 
\scaling is only applied to voting rewards, not to the leader's bonus.
Additionally, combining \scaling and \window, \scaling is applied separately to the part of the reward which is also received on late inclusion and the part, which is only received on timely inclusion.
For example, assume only 90\% of a committee's votes are directly included and 10\% of the votes are delayed and included later. 
Further assume that 30\% of the voters reward is given for timely inclusion. 
Thus, validators with delayed votes will only receive 70\% of their reward. 
For the validators with timely votes, the 30\% received for timely voting will be scaled by 90\% inclusion. 
Thus, these validators will receive $27\%+70\%=97\%$ of their reward. 
The leader receives 90\% of his bonus.

\subsubsection{Reward function}
\label{sec:eth-reward}

To formally discuss Ethereum's rewarding mechanism using our framework, we define the rewarding function  $R(\delta_l,\delta_i,\pow[i],\powersum)$. 
We use the parameters $\rho$ for the fraction of a voters reward,
received on late inclusion and $b$ for the leader's bonus.

\begin{align}
  \label{eq:ethR}
R(\delta_l,\delta_i,\pow[i],\powersum)
  = \rho\,\pow[i]\,R
  + (1-\rho)\,\delta_i\,\powersum\,\pow[i]\,R
  + \delta_l\,b\,\powersum\,R
\end{align}

The first term is the reward received either on timely or late inclusion. 
Since we assume all votes are included within the window, this term is not scaled.
The second term is the reward received only for timely inclusion and it is scaled by the included power $\powersum$. 
The last term is the leaders bonus.
Note that while \aggregation does not appear as a separate term in~\eqref{eq:ethR}, it affects the probability that a player can perform the vote omission attack.

\noindent
To analyse Ethereum's \cost and \effectiveness, we made the following assumptions:

\begin{itemize}
  \item The probability that \emph{all} $k$ aggregators are
        controlled by $p_i$ is negligible
        \(\bigl(P[i]\bigr)^{k}\approx0\).
  \item We ignore the chance that $p_i$ controls $w{+}1$ consecutive
        proposers, \(\bigl(P[i]\bigr)^{\,w+1}\!\approx0\).
\end{itemize}

Based on $R$, the \effectiveness and \cost of vote omission and vote delay attacks are derived as follows:

\begin{equation}
\label{eq:eth-vo-eff}
\effectiveness(S^l_{i\rightarrow j})
    = p_{agg}(i)\,(\,1-\rho).
\end{equation}

\begin{equation}
\label{eq:eth-cost-omit}
\cost(S^l_{i\rightarrow j})
    = \frac{(1-\rho)P[i] + b}{1-\rho}
\end{equation}

\begin{equation}
\label{eq:eth-eff-delay}
\effectiveness(S^v_{i\rightarrow j})
    = (1-\rho)\,P[j]+b
\end{equation}

\begin{equation}
\label{eq:eth-cost-delay}
\cost(S^v_{i\rightarrow j})
    = \frac{1-\rho}{(1-\rho)P[j] + b}
\end{equation}

Note that if we ignore the aggregation probability, the derived \effectiveness and \cost are consistent with Theorems~\ref{thm:costbalance} and~\ref{thm:effbalance}.

\subsubsection{Attacks in Ethereum}

\begin{figure}[ht]
\includegraphics[width=\linewidth]{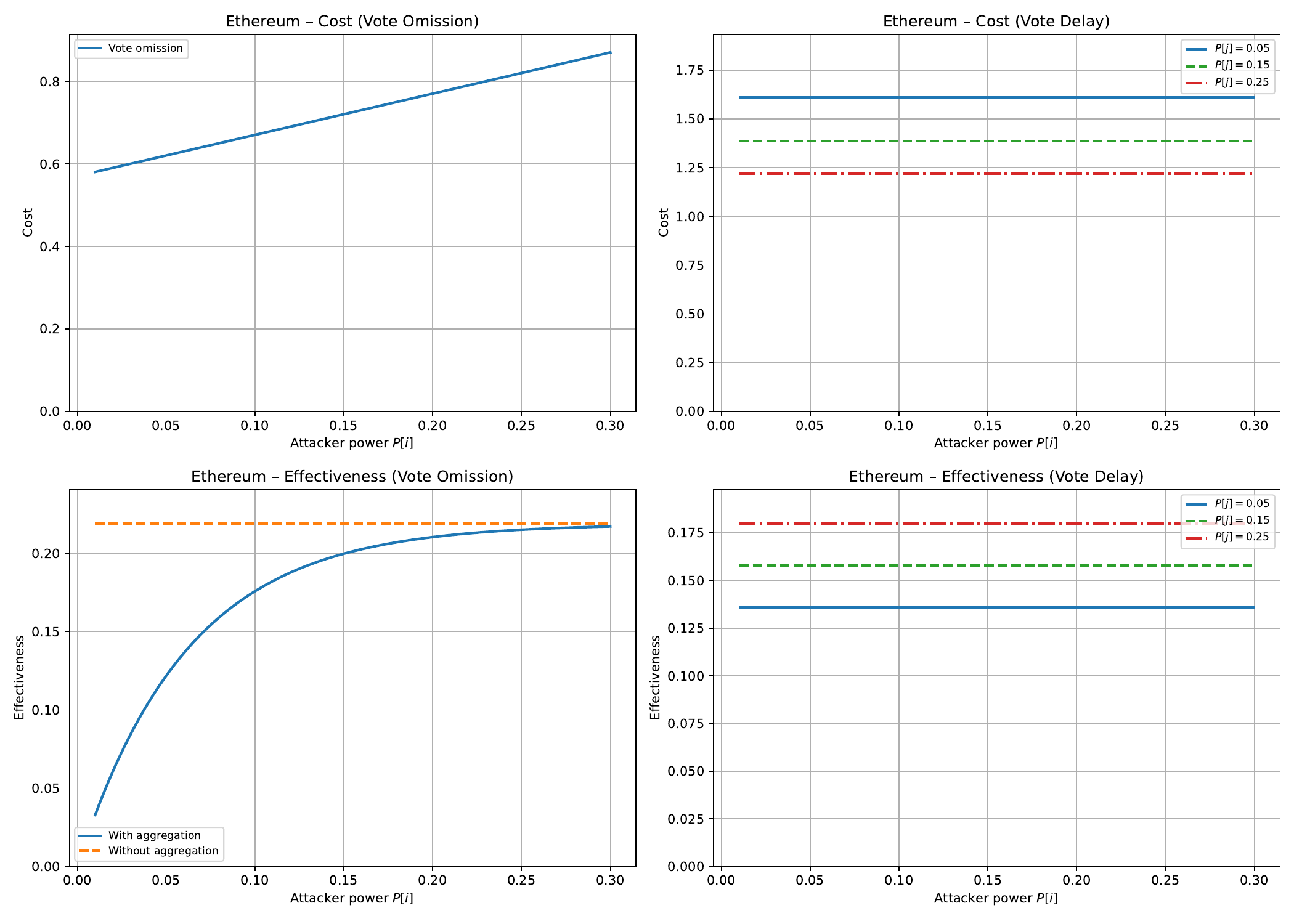}
\caption{
\textbf{Ethereum Attack Metrics versus Attacker Power.}
Top-row panels plot the \cost of vote omission and vote delay as a function of attacker power ($P[i]$) for three victim stakes ($P[j]\in\{0.05,0.15,0.25\}$).
Bottom-row panels show the corresponding \effectiveness curves. 
Parameters: $\rho=0.781$, $b=\tfrac18$.
}
\label{fig:eth-analysis-m}
\end{figure}

As shown in Figure~\ref{fig:eth-analysis-m}, unlike Cosmos, in Ethereum the correct strategy is always a Nash equilibrium. 
In addition, Ethereum manages to achieve a relatively good balance in both attacks compared to Cosmos.  
However, even in Ethereum it is less costly to perform vote omission compared to vote delay. 
For example, \effectiveness of the vote omission attack is $0.2$ for an attacker holding 15\% of the total stake, which means any victim loses 3\% of their total reward. 

\Cost and \effectiveness of vote omission in Ethereum have two interesting properties:
\begin{itemize}
    \item The vote omission attack gets more effective if the attacker power increases. However, it gets more costly as well. 
    \item Both \effectiveness and \cost of the vote omission are independent of the victim power. 
    \item By using \aggregation, Ethereum significantly reduces \effectiveness of vote omission attack for an attacker with a small size. However, for attacker with a larger power, \aggregation does not have any effect. 
\end{itemize}

\subsubsection{Alternative parameters}

\paragraph{Bonus fraction ($b$)}

Increasing $b$ in Ethereum leads into higher \cost in vote omission while lowering down the \cost of vote delay. 
The current value of $b=0.05$ is therefore not balanced, and keeps the cost of both attacks in the range of $[0.2 , 1.8]$. 

\begin{figure*}[ht]
\centering

\begin{subfigure}{0.48\linewidth}
\centering
\includegraphics[width=\linewidth]{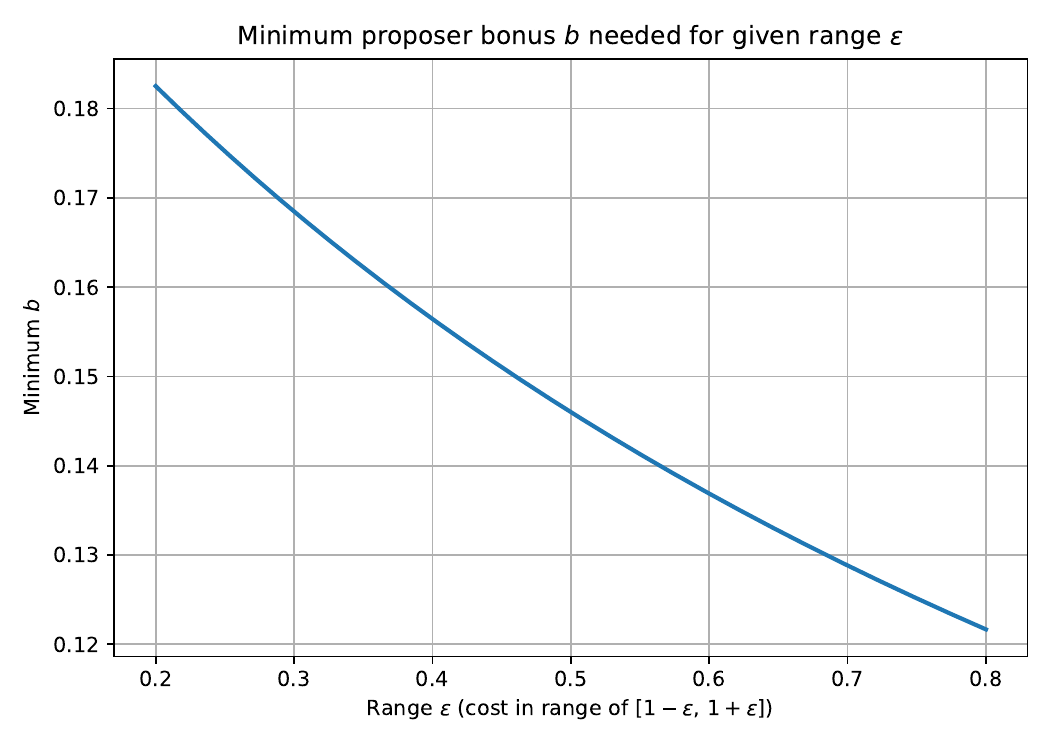}
\caption{\textbf{Minimum bonus derived for different values of $\epsilon$.}}
\label{fig:ethereum-epsilon}
\end{subfigure}
\hfill
\begin{subfigure}{0.48\linewidth}
\centering
\includegraphics[width=\linewidth]{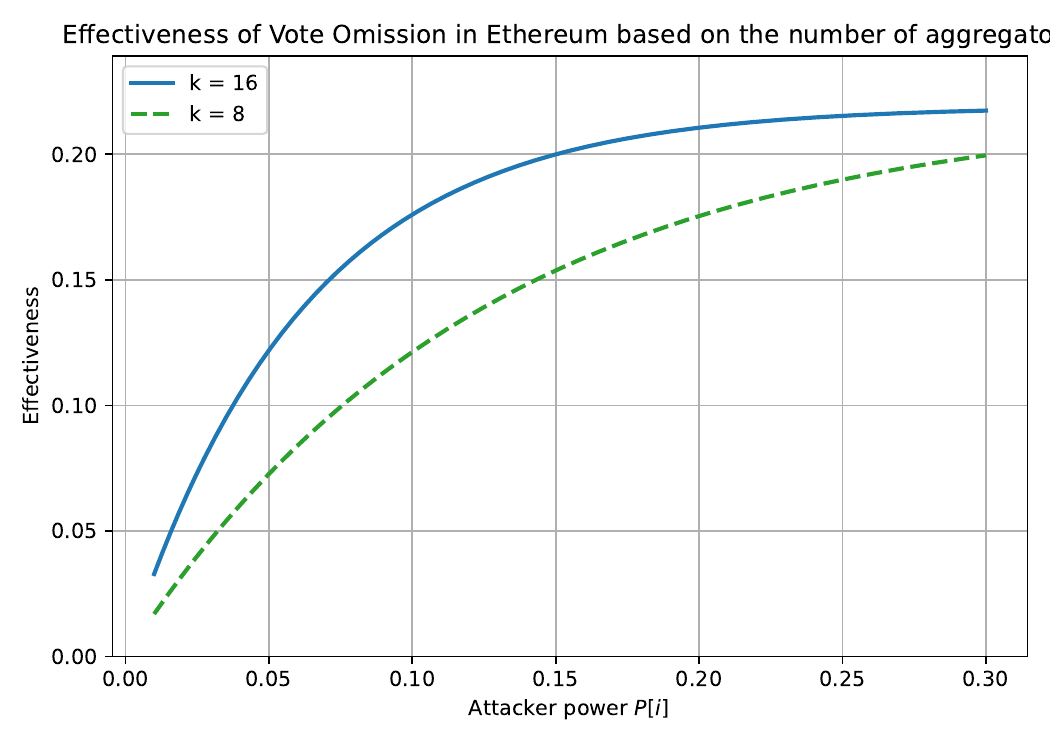}
\caption{\textbf{Ethereum effectiveness based on different number of aggregators ($k = [8,16]$).}}
\label{fig:ethereum-k}
\end{subfigure}

\caption{Comparison of Ethereum reward dynamics under varying parameters.}
\label{fig:ethereum-combined}
\end{figure*}

To find a more balanced $b$, we analyzed different values to find a minimum $b$ that keeps the \cost of both attacks in the range of $[1-\epsilon , 1 + \epsilon]$. 
Figure~\ref{fig:ethereum-epsilon} shows the derived values based on $\epsilon$. 
Note that a larger $b$ is required for a smaller range. 

\begin{corollary}
\label{cor:balanced-b}
Numerical analysis shows that in the Ethereum model, there does not exist a single $b$ that keeps the \cost for both vote omission and vote delay attacks in the range of $[1-\epsilon , 1 + \epsilon]$ if $\epsilon < 0.2$. 
However, the minimum bonus value $b^ * \approx 0.183$ guarantees that for almost all attacker and victim powers $\pow[i],\pow[j]$ the \cost for both vote omission and vote delay attacks is in the range of $[0.8 , 1.2]$.
\end{corollary}

\paragraph{Number of aggregators ($k$)}

Instead of making \cost more balanced by increasing $b$, we can decrease the number of aggregators in each sub-committee ($k$) and make the vote omission attack less likely. 
According to Figure~\ref{fig:ethereum-k}, \effectiveness in Ethereum is significantly reduced if we have $8$ aggregators instead of $16$.
While even smaller number of aggregators may still be beneficial, one can no longer ignore the possibility that one player controls all aggregators.

\section{Related Work}
\label{sec:related}
While this paper focuses on \dop attacks, a number of works have analyzed and critiqued the reward mechanisms in Cosmos and Ethereum.
Post-merge Ethereum has been examined extensively, with researchers identifying vulnerabilities and incentive misalignments in its reward design~\cite{cortes2023autopsy, yan2024analyzing, pavloff2024byzantine, pavloff2025incentive}. 
In the case of Cosmos, several studies highlight that following the protocol is not a Nash Equilibrium~\cite{amoussou2018correctness, eiffel, lagaillardie2019computational}, primarily due to vote omission attacks.
Lagaillardie et al.~\cite{lagaillardie2019computational} propose a \window mechanism to mitigate this issue, while Baloochestani et al.~\cite{rebop} introduce Rebop, a reputation-based leader election protocol that reduces the incentive to omit votes.
Their analysis also shows that increasing the cost of vote omission can make vote delay attacks more appealing, illustrating a trade-off that remains unresolved.
To address this, they later propose Iniva~\cite{iniva}, a tree-based vote aggregation protocol with fallback paths.
Iniva combines indivisible multi-signatures with incentive mechanisms to promote participation and reduce the effectiveness of vote omission attacks.

Beyond Cosmos and Ethereum, other blockchains have also incorporated mechanisms that relate to \dop attack resilience. 
In Harmony~\cite{harmony}, rewards are distributed among all validators whose votes are included in the blockchain. 
Harmony also uses a redistribution model similar to Cosmos where the rewards of the absent processes are redistributed among others. 
Hemati et al.~\cite{hemati2021incentive} show that this mechanism is prone to the vote omission attack, and the leader is incentivized to omit some votes to increase its own share. 
Therefore, Avarikioti et al.~\cite{avarikioti2022harmony} proposed Harmony to use a superlinear version of \scaling to discourage malicious behaviour. 
Tezos~\cite{goodman2014tezos} adopts the \textit{bonus threshold} mechanism and encourages inclusion of votes by paying a bonus for extra attestations included by the leaders beyond a certain threshold~\cite{opentezos_rewards}. 

Finally, many other systems implicitly expose themselves to \dop attacks by rewarding participants based on their contributions. 
For example, in Polkadot~\cite{wood2016polkadot}, participants earn era points for submitting validity statements for blocks on side-chains.
In Solana~\cite{yakovenko2018solana}, votes from validators are included as signatures and validators receive smaller rewards if their votes are included later.
Sui and IOTA also reward validators according to their performance and contribution.

\section{Conclusion}
\label{sec:conclusion}
This paper introduced a game-theoretical framework and metrics to study \dop attacks, where attackers reduce others' profits even at a cost to themselves.
As a concrete case, we modeled vote collection in committee-based blockchains as a game and proposed metrics to quantify both attacker cost and victim loss.
Using this model, we analyzed the reward mechanisms of Ethereum and Cosmos, and showed how imbalances in these systems can make honest behavior suboptimal.

Our results show that even when protection mechanisms are in place, they are often incomplete.
For example, in Cosmos, vote omission attacks are actually profitable, while Ethereum offers stronger, but still not fully balanced, protections.
Our framework enables protocol designers to fine-tune parameters and identify reward configurations that achieve better incentive alignment.

\bibliography{references}

\end{document}